\documentclass[sigconf]{acmart} 

\AtBeginDocument{%
  \providecommand\BibTeX{{%
    \normalfont B\kern-0.5em{\scshape i\kern-0.25em b}\kern-0.8em\TeX}}}

\copyrightyear{2021}
\acmYear{2021}
\setcopyright{acmcopyright}
\acmConference[Koli Calling '21]{21st Koli Calling International Conference on Computing Education Research}{November 18--21, 2021}{Joensuu, Finland}
\acmBooktitle{21st Koli Calling International Conference on Computing Education Research (Koli Calling '21), November 18--21, 2021, Joensuu, Finland}
\acmPrice{15.00}
\acmDOI{10.1145/3488042.3488048}
\acmISBN{978-1-4503-8488-9/21/11}

\usepackage{xspace}
\usepackage{tabularx}
\usepackage{colortbl}
\usepackage{subfig}
\usepackage{multirow}
\usepackage{booktabs}
\usepackage{listings}
\usepackage{xcolor}
\usepackage{siunitx}
\usepackage[capitalise]{cleveref}
\usepackage{fp}
\usepackage{collcell}
\usepackage{amsmath}
\usepackage{paralist}
\usepackage{circledsteps}
\usepackage{tikz}

\newcommand\definetool[2]{\newcommand{#1}{{\textsc{#2}}\xspace}}
\definetool{\Scratch}{Scratch}
\definetool{\leila}{LeILa}
\definetool{\whisker}{Whisker}
\definetool{\litterbox}{LitterBox}
\definetool{\bastet}{Bastet}
\definetool{\scratchblocks}{scratchblocks}
\definetool{\Purplemash}{Purple Mash}
\definetool{\Codeorg}{Code.org}

\colorlet{punct}{red!60!black}
\definecolor{background}{HTML}{EEEEEE}
\definecolor{delim}{RGB}{20,105,176}
\colorlet{numb}{magenta!60!black}

\definecolor{Gray}{gray}{0.9}

\lstdefinelanguage{json}{
    basicstyle=\normalfont\ttfamily,
    numbers=left,
    numberstyle=\scriptsize,
    stepnumber=1,
    numbersep=8pt,
    showstringspaces=false,
    breaklines=true,
    frame=lines,
    backgroundcolor=\color{background},
    literate=
     *{0}{{{\color{numb}0}}}{1}
      {1}{{{\color{numb}1}}}{1}
      {2}{{{\color{numb}2}}}{1}
      {3}{{{\color{numb}3}}}{1}
      {4}{{{\color{numb}4}}}{1}
      {5}{{{\color{numb}5}}}{1}
      {6}{{{\color{numb}6}}}{1}
      {7}{{{\color{numb}7}}}{1}
      {8}{{{\color{numb}8}}}{1}
      {9}{{{\color{numb}9}}}{1}
      {:}{{{\color{punct}{:}}}}{1}
      {,}{{{\color{punct}{,}}}}{1}
      {\{}{{{\color{delim}{\{}}}}{1}
      {\}}{{{\color{delim}{\}}}}}{1}
      {[}{{{\color{delim}{[}}}}{1}
      {]}{{{\color{delim}{]}}}}{1},
}

\usepackage{fbox}
\newcommand{\summary}[2]{
        \vspace{2mm}
        \noindent
        \fbox{%
            \parbox{.97\linewidth}{%
                    \textbf{#1 Summary.}
                #2
            }%
        }%
}%

\usepackage[colorinlistoftodos,prependcaption,textsize=tiny]{todonotes}

\begin{document}

\newcommand{\PValueOpportunitiesSkillsacquisition}{0.000}
\newcommand{\PercentPracticeOpportunitiesSkillsacquisition}{0.55}
\newcommand{\PercentTeacherOpportunitiesSkillsacquisition}{0.814433}
\newcommand{\PValueOpportunitiesTransferableknowledge}{0.8702298}
\newcommand{\PercentPracticeOpportunitiesTransferableknowledge}{0.345}
\newcommand{\PercentTeacherOpportunitiesTransferableknowledge}{0.3402062}
\newcommand{\PValueOpportunitiesSocialimportance}{0.000}
\newcommand{\PercentPracticeOpportunitiesSocialimportance}{0.1}
\newcommand{\PercentTeacherOpportunitiesSocialimportance}{0.2886598}
\newcommand{\PValueOpportunitiesEarlysupport}{0.000}
\newcommand{\PercentPracticeOpportunitiesEarlysupport}{0.07}
\newcommand{\PercentTeacherOpportunitiesEarlysupport}{0.2989691}
\newcommand{\PValueOpportunitiesCrosscurricularlearning}{0.02842621}
\newcommand{\PercentPracticeOpportunitiesCrosscurricularlearning}{0.105}
\newcommand{\PercentTeacherOpportunitiesCrosscurricularlearning}{0.03092784}
\newcommand{\PValueOpportunitiesMethods}{0.000}
\newcommand{\PercentPracticeOpportunitiesMethods}{0.035}
\newcommand{\PercentTeacherOpportunitiesMethods}{0.1649485}
\newcommand{\PValueOpportunitiesDiversity}{0.126062}
\newcommand{\PercentPracticeOpportunitiesDiversity}{0.055}
\newcommand{\PercentTeacherOpportunitiesDiversity}{0.1030928}
\newcommand{\PValueOpportunitiesSkillsacquisitionholistic}{0.000}
\newcommand{\PercentPracticeOpportunitiesSkillsacquisitionholistic}{0.265}
\newcommand{\PercentTeacherOpportunitiesSkillsacquisitionholistic}{0.5979381}
\newcommand{\PValueOpportunitiesSkillsacquisitionaffective}{0.03947107}
\newcommand{\PercentPracticeOpportunitiesSkillsacquisitionaffective}{0.23}
\newcommand{\PercentTeacherOpportunitiesSkillsacquisitionaffective}{0.3298969}
\newcommand{\PValueOpportunitiesSkillsacquisitioncognitive}{0.1541637}
\newcommand{\PercentPracticeOpportunitiesSkillsacquisitioncognitive}{0.155}
\newcommand{\PercentTeacherOpportunitiesSkillsacquisitioncognitive}{0.09278351}
\newcommand{\PValueOpportunitiesSkillsacquisitiongeneral}{0.03751223}
\newcommand{\PercentPracticeOpportunitiesSkillsacquisitiongeneral}{0.065}
\newcommand{\PercentTeacherOpportunitiesSkillsacquisitiongeneral}{0.01030928}
\newcommand{\PValueOpportunitiesSkillsacquisitionmetacognitiveselfreliance}{0.009829577}
\newcommand{\PercentPracticeOpportunitiesSkillsacquisitionmetacognitiveselfreliance}{0.01}
\newcommand{\PercentTeacherOpportunitiesSkillsacquisitionmetacognitiveselfreliance}{0.06185567}
\newcommand{\PValueOpportunitiesSkillsacquisitionholisticgeneral}{0.03437452}
\newcommand{\PercentPracticeOpportunitiesSkillsacquisitionholisticgeneral}{0.045}
\newcommand{\PercentTeacherOpportunitiesSkillsacquisitionholisticgeneral}{0}
\newcommand{\PValueOpportunitiesSkillsacquisitionholisticcreativity}{0.6126122}
\newcommand{\PercentPracticeOpportunitiesSkillsacquisitionholisticcreativity}{0.085}
\newcommand{\PercentTeacherOpportunitiesSkillsacquisitionholisticcreativity}{0.1030928}
\newcommand{\PValueOpportunitiesSkillsacquisitionholisticdigitalliteracy}{0.000}
\newcommand{\PercentPracticeOpportunitiesSkillsacquisitionholisticdigitalliteracy}{0.06}
\newcommand{\PercentTeacherOpportunitiesSkillsacquisitionholisticdigitalliteracy}{0.443299}
\newcommand{\PValueOpportunitiesSkillsacquisitionholisticcomputationalliteracy}{0.000}
\newcommand{\PercentPracticeOpportunitiesSkillsacquisitionholisticcomputationalliteracy}{0.115}
\newcommand{\PercentTeacherOpportunitiesSkillsacquisitionholisticcomputationalliteracy}{0.2886598}
\newcommand{\PValueOpportunitiesSkillsacquisitionaffectivegeneral}{0.007666266}
\newcommand{\PercentPracticeOpportunitiesSkillsacquisitionaffectivegeneral}{0.005}
\newcommand{\PercentTeacherOpportunitiesSkillsacquisitionaffectivegeneral}{0.05154639}
\newcommand{\PValueOpportunitiesSkillsacquisitionaffectivetalent}{0.000}
\newcommand{\PercentPracticeOpportunitiesSkillsacquisitionaffectivetalent}{0}
\newcommand{\PercentTeacherOpportunitiesSkillsacquisitionaffectivetalent}{0.07216495}
\newcommand{\PValueOpportunitiesSkillsacquisitionaffectivefun}{0.1065161}
\newcommand{\PercentPracticeOpportunitiesSkillsacquisitionaffectivefun}{0.045}
\newcommand{\PercentTeacherOpportunitiesSkillsacquisitionaffectivefun}{0.09278351}
\newcommand{\PValueOpportunitiesSkillsacquisitionaffectiveselfconfidence}{0.1343121}
\newcommand{\PercentPracticeOpportunitiesSkillsacquisitionaffectiveselfconfidence}{0.06}
\newcommand{\PercentTeacherOpportunitiesSkillsacquisitionaffectiveselfconfidence}{0.02061856}
\newcommand{\PValueOpportunitiesSkillsacquisitionaffectivereductionofprejudices}{0.4857515}
\newcommand{\PercentPracticeOpportunitiesSkillsacquisitionaffectivereductionofprejudices}{0.085}
\newcommand{\PercentTeacherOpportunitiesSkillsacquisitionaffectivereductionofprejudices}{0.06185567}
\newcommand{\PValueOpportunitiesSkillsacquisitionaffectiveinterestandmotivation}{0.0222575}
\newcommand{\PercentPracticeOpportunitiesSkillsacquisitionaffectiveinterestandmotivation}{0.085}
\newcommand{\PercentTeacherOpportunitiesSkillsacquisitionaffectiveinterestandmotivation}{0.1752577}
\newcommand{\PValueOpportunitiesSkillsacquisitioncognitivegeneral}{0.1904487}
\newcommand{\PercentPracticeOpportunitiesSkillsacquisitioncognitivegeneral}{0.01}
\newcommand{\PercentTeacherOpportunitiesSkillsacquisitioncognitivegeneral}{0.03092784}
\newcommand{\PValueOpportunitiesSkillsacquisitioncognitivelinguisticstimulation}{0.5454984}
\newcommand{\PercentPracticeOpportunitiesSkillsacquisitioncognitivelinguisticstimulation}{0.02}
\newcommand{\PercentTeacherOpportunitiesSkillsacquisitioncognitivelinguisticstimulation}{0.01030928}
\newcommand{\PValueOpportunitiesSkillsacquisitioncognitiveproblemsolvingandlogicalthinking}{0.08984134}
\newcommand{\PercentPracticeOpportunitiesSkillsacquisitioncognitiveproblemsolvingandlogicalthinking}{0.14}
\newcommand{\PercentTeacherOpportunitiesSkillsacquisitioncognitiveproblemsolvingandlogicalthinking}{0.07216495}
\newcommand{\PValueOpportunitiesTransferableknowledgegeneralfuture}{0.004549165}
\newcommand{\PercentPracticeOpportunitiesTransferableknowledgegeneralfuture}{0.075}
\newcommand{\PercentTeacherOpportunitiesTransferableknowledgegeneralfuture}{0.185567}
\newcommand{\PValueOpportunitiesTransferableknowledgeprofessionallife}{0.05737969}
\newcommand{\PercentPracticeOpportunitiesTransferableknowledgeprofessionallife}{0.15}
\newcommand{\PercentTeacherOpportunitiesTransferableknowledgeprofessionallife}{0.07216495}
\newcommand{\PValueOpportunitiesTransferableknowledgesecondaryschools}{0.02607864}
\newcommand{\PercentPracticeOpportunitiesTransferableknowledgesecondaryschools}{0.065}
\newcommand{\PercentTeacherOpportunitiesTransferableknowledgesecondaryschools}{0.1443299}
\newcommand{\PValueOpportunitiesTransferableknowledgegeneral}{0.005754305}
\newcommand{\PercentPracticeOpportunitiesTransferableknowledgegeneral}{0.075}
\newcommand{\PercentTeacherOpportunitiesTransferableknowledgegeneral}{0}
\newcommand{\PValueOpportunitiesSocialimportancestudentsslifeworld}{0.003940459}
\newcommand{\PercentPracticeOpportunitiesSocialimportancestudentsslifeworld}{0}
\newcommand{\PercentTeacherOpportunitiesSocialimportancestudentsslifeworld}{0.04123711}
\newcommand{\PValueOpportunitiesSocialimportanceincreasingdigitization}{0.000}
\newcommand{\PercentPracticeOpportunitiesSocialimportanceincreasingdigitization}{0.1}
\newcommand{\PercentTeacherOpportunitiesSocialimportanceincreasingdigitization}{0.2680412}
\newcommand{\PValueOpportunitiesEarlysupportgeneral}{0.000}
\newcommand{\PercentPracticeOpportunitiesEarlysupportgeneral}{0.07}
\newcommand{\PercentTeacherOpportunitiesEarlysupportgeneral}{0.2989691}
\newcommand{\PValueOpportunitiesCrosscurricularlearninggeneral}{0.02842621}
\newcommand{\PercentPracticeOpportunitiesCrosscurricularlearninggeneral}{0.105}
\newcommand{\PercentTeacherOpportunitiesCrosscurricularlearninggeneral}{0.03092784}
\newcommand{\PValueOpportunitiesMethodsvariety}{0.0127253}
\newcommand{\PercentPracticeOpportunitiesMethodsvariety}{0}
\newcommand{\PercentTeacherOpportunitiesMethodsvariety}{0.03092784}
\newcommand{\PValueOpportunitiesMethodsexploringdiscoverylearning}{0.05354033}
\newcommand{\PercentPracticeOpportunitiesMethodsexploringdiscoverylearning}{0.025}
\newcommand{\PercentTeacherOpportunitiesMethodsexploringdiscoverylearning}{0.07216495}
\newcommand{\PValueOpportunitiesMethodsplayfulchildorientedrealization}{0.02759516}
\newcommand{\PercentPracticeOpportunitiesMethodsplayfulchildorientedrealization}{0.015}
\newcommand{\PercentTeacherOpportunitiesMethodsplayfulchildorientedrealization}{0.06185567}
\newcommand{\PValueOpportunitiesDiversityalignmentofstudents}{0.500003}
\newcommand{\PercentPracticeOpportunitiesDiversityalignmentofstudents}{0.035}
\newcommand{\PercentTeacherOpportunitiesDiversityalignmentofstudents}{0.02061856}
\newcommand{\PValueOpportunitiesDiversitygirls}{0.02598274}
\newcommand{\PercentPracticeOpportunitiesDiversitygirls}{0.02}
\newcommand{\PercentTeacherOpportunitiesDiversitygirls}{0.07216495}
\newcommand{\PValueOpportunitiesDiversitybothgirlsandboys}{0.04245361}
\newcommand{\PercentPracticeOpportunitiesDiversitybothgirlsandboys}{0}
\newcommand{\PercentTeacherOpportunitiesDiversitybothgirlsandboys}{0.02061856}
\newcommand{\PValueChallengesGovernment}{0.4707442}
\newcommand{\PercentPracticeChallengesGovernment}{0.17}
\newcommand{\PercentTeacherChallengesGovernment}{0.1340206}
\newcommand{\PValueChallengesParents}{0.01558355}
\newcommand{\PercentPracticeChallengesParents}{0.04}
\newcommand{\PercentTeacherChallengesParents}{0.1134021}
\newcommand{\PValueChallengesProgramming}{0.0717173}
\newcommand{\PercentPracticeChallengesProgramming}{0.215}
\newcommand{\PercentTeacherChallengesProgramming}{0.3092784}
\newcommand{\PValueChallengesSchool}{0.000}
\newcommand{\PercentPracticeChallengesSchool}{0.45}
\newcommand{\PercentTeacherChallengesSchool}{0.1958763}
\newcommand{\PValueChallengesStudents}{0.000}
\newcommand{\PercentPracticeChallengesStudents}{0.415}
\newcommand{\PercentTeacherChallengesStudents}{0.6391753}
\newcommand{\PValueChallengesTeachers}{0.1447181}
\newcommand{\PercentPracticeChallengesTeachers}{0.365}
\newcommand{\PercentTeacherChallengesTeachers}{0.4329897}
\newcommand{\PValueChallengesGovernmentorganisational}{0.4707442}
\newcommand{\PercentPracticeChallengesGovernmentorganisational}{0.17}
\newcommand{\PercentTeacherChallengesGovernmentorganisational}{0.1340206}
\newcommand{\PValueChallengesParentsaffective}{0.02731743}
\newcommand{\PercentPracticeChallengesParentsaffective}{0.01}
\newcommand{\PercentTeacherChallengesParentsaffective}{0.05154639}
\newcommand{\PValueChallengesParentsorganisational}{0.1925003}
\newcommand{\PercentPracticeChallengesParentsorganisational}{0.03}
\newcommand{\PercentTeacherChallengesParentsorganisational}{0.06185567}
\newcommand{\PValueChallengesProgrammingdebuggingandproblems}{0.02560878}
\newcommand{\PercentPracticeChallengesProgrammingdebuggingandproblems}{0.09}
\newcommand{\PercentTeacherChallengesProgrammingdebuggingandproblems}{0.02061856}
\newcommand{\PValueChallengesProgrammingcomplexity}{0.000}
\newcommand{\PercentPracticeChallengesProgrammingcomplexity}{0.13}
\newcommand{\PercentTeacherChallengesProgrammingcomplexity}{0.2886598}
\newcommand{\PValueChallengesSchoolorganisational}{0.000}
\newcommand{\PercentPracticeChallengesSchoolorganisational}{0.45}
\newcommand{\PercentTeacherChallengesSchoolorganisational}{0.1958763}
\newcommand{\PValueChallengesStudentsaffective}{0.9351005}
\newcommand{\PercentPracticeChallengesStudentsaffective}{0.1}
\newcommand{\PercentTeacherChallengesStudentsaffective}{0.1030928}
\newcommand{\PValueChallengesStudentscognitive}{0.000}
\newcommand{\PercentPracticeChallengesStudentscognitive}{0.24}
\newcommand{\PercentTeacherChallengesStudentscognitive}{0.4742268}
\newcommand{\PValueChallengesStudentsheterogeneity}{0.002479414}
\newcommand{\PercentPracticeChallengesStudentsheterogeneity}{0.065}
\newcommand{\PercentTeacherChallengesStudentsheterogeneity}{0.1752577}
\newcommand{\PValueChallengesStudentsyoungage}{0.500003}
\newcommand{\PercentPracticeChallengesStudentsyoungage}{0.035}
\newcommand{\PercentTeacherChallengesStudentsyoungage}{0.02061856}
\newcommand{\PValueChallengesStudentsmetacognitive}{0.3928961}
\newcommand{\PercentPracticeChallengesStudentsmetacognitive}{0.065}
\newcommand{\PercentTeacherChallengesStudentsmetacognitive}{0.09278351}
\newcommand{\PValueChallengesTeachersaffective}{0.2735912}
\newcommand{\PercentPracticeChallengesTeachersaffective}{0.075}
\newcommand{\PercentTeacherChallengesTeachersaffective}{0.1134021}
\newcommand{\PValueChallengesTeacherscognitive}{0.1406709}
\newcommand{\PercentPracticeChallengesTeacherscognitive}{0.245}
\newcommand{\PercentTeacherChallengesTeacherscognitive}{0.1649485}
\newcommand{\PValueChallengesTeachersmethodicalanddidactic}{0.000}
\newcommand{\PercentPracticeChallengesTeachersmethodicalanddidactic}{0.11}
\newcommand{\PercentTeacherChallengesTeachersmethodicalanddidactic}{0.2680412}
\newcommand{\StrategyPercentStrategiesforChallengesParents}{0.105}
\newcommand{\StrategyPercentStrategiesforChallengesSociety}{0.01}
\newcommand{\StrategyPercentStrategiesforChallengesGovernment}{0.415}
\newcommand{\StrategyPercentStrategiesforChallengesTeachers}{0.965}
\newcommand{\StrategyPercentStrategiesforChallengesSchool}{0.5}
\newcommand{\StrategyPercentAnalysisToolsadvantages}{0.185}
\newcommand{\StrategyPercentAnalysisToolsaims}{0.14}
\newcommand{\StrategyPercentAnalysisToolsalternatives}{0.115}
\newcommand{\StrategyPercentAnalysisToolsdisadvantages}{0.12}
\newcommand{\StrategyPercentAnalysisToolstargetgroup}{0.415}
\newcommand{\StrategyPercentAnalysisToolsuncertainty}{0.33}
\newcommand{\StrategyPercentAnalysisToolsusecases}{0.325}
\newcommand{\StrategyPercentStrategiesforChallengesParentsContent}{0.06}
\newcommand{\StrategyPercentStrategiesforChallengesParentsSupport}{0.03}
\newcommand{\StrategyPercentStrategiesforChallengesParentsMeetings}{0.04}
\newcommand{\StrategyPercentStrategiesforChallengesSocietyrethinksociologically}{0.01}
\newcommand{\StrategyPercentStrategiesforChallengesGovernmentreorganisecurricula}{0.23}
\newcommand{\StrategyPercentStrategiesforChallengesGovernmentmorefundsforCSE}{0.265}
\newcommand{\StrategyPercentStrategiesforChallengesTeachersparticipateintraining}{0.68}
\newcommand{\StrategyPercentStrategiesforChallengesTeachersteaching}{0.425}
\newcommand{\StrategyPercentStrategiesforChallengesTeacherscontentoftraining}{0.235}
\newcommand{\StrategyPercentStrategiesforChallengesTeacherscooperatewithothers}{0.155}
\newcommand{\StrategyPercentStrategiesforChallengesTeacherstime}{0.14}
\newcommand{\StrategyPercentStrategiesforChallengesTeachersgetsupportthroughmaterial}{0.105}
\newcommand{\StrategyPercentStrategiesforChallengesSchooljointapproach}{0.04}
\newcommand{\StrategyPercentStrategiesforChallengesSchoolcourseafterschool}{0.045}
\newcommand{\StrategyPercentStrategiesforChallengesSchoolmediaequipment}{0.435}
\newcommand{\StrategyPercentStrategiesforChallengesSchoolresponsibleteacher}{0.03}
\newcommand{\StrategyPercentAnalysisToolsadvantagesdealwithmanystudents}{0.065}
\newcommand{\StrategyPercentAnalysisToolsadvantagesindependenceoftheteacher}{0.045}
\newcommand{\StrategyPercentAnalysisToolsadvantageseasy}{0.035}
\newcommand{\StrategyPercentAnalysisToolsadvantagesimmediate}{0.035}
\newcommand{\StrategyPercentAnalysisToolsadvantagesmoreconfidenceofteachers}{0.03}
\newcommand{\StrategyPercentAnalysisToolsadvantagescontinuous}{0.015}
\newcommand{\StrategyPercentAnalysisToolsaimstime}{0.12}
\newcommand{\StrategyPercentAnalysisToolsaimsstudentsacquisitionofcompetencies}{0.025}
\newcommand{\StrategyPercentAnalysisToolsalternativeshumanfeedback}{0.08}
\newcommand{\StrategyPercentAnalysisToolsalternativesnotonlytools}{0.045}
\newcommand{\StrategyPercentAnalysisToolsdisadvantagesinsufficient}{0.045}
\newcommand{\StrategyPercentAnalysisToolsdisadvantagestaskoftheteacher}{0.03}
\newcommand{\StrategyPercentAnalysisToolsdisadvantagesnoindividualisation}{0.025}
\newcommand{\StrategyPercentAnalysisToolsdisadvantagescomplex}{0.02}
\newcommand{\StrategyPercentAnalysisToolstargetgroupforteachers}{0.225}
\newcommand{\StrategyPercentAnalysisToolstargetgroupgeneralusefulness}{0.145}
\newcommand{\StrategyPercentAnalysisToolstargetgroupforstudents}{0.075}
\newcommand{\StrategyPercentAnalysisToolsuncertaintynoexperience}{0.295}
\newcommand{\StrategyPercentAnalysisToolsuncertaintynounderstanding}{0.035}
\newcommand{\StrategyPercentAnalysisToolsusecasesfeedback}{0.145}
\newcommand{\StrategyPercentAnalysisToolsusecasesdebugging}{0.085}
\newcommand{\StrategyPercentAnalysisToolsusecasesdiagnosis}{0.075}
\newcommand{\StrategyPercentAnalysisToolsusecasessupportandtendencies}{0.075}
\newcommand{\StrategyPercentAnalysisToolsusecasesassessment}{0.02}
\newcommand{\StrategyPercentStrategiesforOpportunitiesSocialandprivateimportancepracticalapplication}{0.16}
\newcommand{\StrategyPercentStrategiesforOpportunitiesSocialandprivateimportancestudentslifeworld}{0.235}
\newcommand{\StrategyPercentStrategiesforOpportunitiesSocialandprivateimportancerelevanceofprogramming}{0.205}
\newcommand{\StrategyPercentStrategiesforOpportunitiesSocialandprivateimportancefuturejobopportunities}{0.285}
\newcommand{\StrategyPercentStrategiesforOpportunitiesMethodsactivitieschallengeshomework}{0.195}
\newcommand{\StrategyPercentStrategiesforOpportunitiesMethodsgeneral}{0.33}
\newcommand{\StrategyPercentStrategiesforOpportunitiesMethodsindividualneedsfeedbackteamworklearningstrategies}{0.115}
\newcommand{\StrategyPercentStrategiesforOpportunitiesMethodsprograms}{0.045}
\newcommand{\StrategyPercentStrategiesforOpportunitiesAffectivestimulatingalternativewaysofthinking}{0.07}
\newcommand{\StrategyPercentStrategiesforOpportunitiesAffectivestimulatingstudentscreativity}{0.14}
\newcommand{\StrategyPercentStrategiesforOpportunitiesAffectivemindsetandintegration}{0.13}
\newcommand{\StrategyPercentStrategiesforOpportunitiesAffectiveencouragementandenthusiam}{0.135}
\newcommand{\StrategyPercentStrategiesforOpportunitiesAffectiveimprovingaffectivefactorsingeneral}{0.11}
\newcommand{\StrategyPercentStrategiesforOpportunitiesOrganizationalinvitingexperts}{0.06}
\newcommand{\StrategyPercentStrategiesforOpportunitiesOrganizationalrequirementsliketrainingandequipment}{0.27}
\newcommand{\StrategyPercentStrategiesforOpportunitiesGainingprogrammingskillsgeneral}{0.225}
\newcommand{\StrategyPercentStrategiesforOpportunitiesCrosscurricularlearninggeneral}{0.21}
\newcommand{\StrategyPercentStrategiesforOpportunitiesEarlysupportgeneral}{0.105}
\newcommand{\StrategyPercentHomogeneousCoursesRatingnegative}{0.27}
\newcommand{\StrategyPercentHomogeneousCoursesRatingneutral}{0.275}
\newcommand{\StrategyPercentHomogeneousCoursesRatingnotdoneyet}{0.04}
\newcommand{\StrategyPercentHomogeneousCoursesRatingpositive}{0.345}
\newcommand{\StrategyPercentHomogeneousCoursesRatingunsure}{0.085}
\newcommand{\StrategyPercentHomogeneousCoursesDifferencestopics}{0.01}
\newcommand{\StrategyPercentHomogeneousCoursesDifferencesaffective}{0.01}
\newcommand{\StrategyPercentHomogeneousCoursesDifferencesgeneral}{0.02}
\newcommand{\StrategyPercentHomogeneousCoursesSimilaritiescognitive}{0.035}
\newcommand{\StrategyPercentHomogeneousCoursesSimilaritiesaffective}{0.075}
\newcommand{\StrategyPercentHomogeneousCoursesSimilaritiesperformance}{0.02}
\newcommand{\StrategyPercentHomogeneousCoursesSimilaritiesgeneral}{0.055}
\newcommand{\ChallengePercentOne}{0.105}
\newcommand{\ChallengePercentTwo}{0.21}
\newcommand{\ChallengePercentThree}{0.535}
\newcommand{\ChallengePercentFour}{0.305}
\newcommand{\ChallengePercentFive}{0.205}
\newcommand{\ChallengePercentSix}{0.695}
\newcommand{\ChallengePercentSeven}{0.035}
\newcommand{\ChallengePercentEight}{0.2}
\newcommand{\ChallengePercentNine}{0.115}
\newcommand{\ChallengePercentTen}{0.355}
\newcommand{\OpportunityPercentOne}{0.545}
\newcommand{\OpportunityPercentTwo}{0.465}
\newcommand{\OpportunityPercentThree}{0.65}
\newcommand{\OpportunityPercentFour}{0.26}
\newcommand{\OpportunityPercentFive}{0.085}
\newcommand{\OpportunityPercentSix}{0.07}
\newcommand{\OpportunityPercentSeven}{0.125}
\newcommand{\OpportunityPercentEight}{0.4}
\newcommand{\OpportunityPercentNine}{0.24}
\newcommand{\OpportunityPercentTen}{0.155}

\newcommand{\numParticipantsTraining}{97~}
\newcommand{\numParticipantsTrainingFemale}{89~}
\newcommand{\numParticipantsTrainingMale}{8~}
\newcommand{\ageTrainingMedian}{20~}
\newcommand{\percentageProgrammingExperience}{56}
\newcommand{\numParticipantsTrainingTotal}{242~}
\renewcommand{\tt}{TT\xspace}
\newcommand{\tp}{TP\xspace}
\newcommand{\ttEnd}{TT}
\newcommand{\tpEnd}{TP}
\newcommand{\didacticconsiderations}{the teaching approach~}
\newcommand{\irr}{0.69}

\newlength\MAX  \setlength\MAX{5mm}
\newcommand*\Chart[1]{\rlap{\textcolor{black!20}{\rule{\MAX}{2ex}}}\rule{#1\MAX}{2ex}}

\newcommand{\numParticipantsPractice}{200~}

\newcommand{\pvalue}[1]{%
  \ifthenelse{\equal{#1}{0.000}}%
    {$p < 0.001$}%
    {$p = \num[round-mode=places,round-precision=3]{#1}$}%
}

\newcommand{\tpquote}[2]{%
\noindent\Circled{\textsf{\tiny TP #1}}~``\textit{#2}''%
}

\newcommand{\ttquote}[2]{%
\noindent\Circled{\textsf{\tiny TT #1}}~``\textit{#2}''%
}

\newcommand\printpercent[1]{\FPeval\result{round(#1*100,1)}\num[round-mode=places,round-precision=1]{\result}~\%}

\newcommand{\DrawPercentageBar}[1]{%
  \begin{tikzpicture}
    \fill[color=black]   (0.0 , 0.0) rectangle (#1*5ex , 1.7ex );
    \fill[color=lightgray] (#1*5ex  , 0.0) rectangle (5.0ex, 1.7ex);
  \end{tikzpicture}%
}

\newcommand\zz[1]{%
\ifdim#1pt<1pt\cellcolor{white}\else
\ifdim#1pt<5pt\cellcolor{gray!10}\else
\ifdim#1pt<10pt\cellcolor{gray!20}\else
\ifdim#1pt<15pt\cellcolor{gray!30}\else
\ifdim#1pt<20pt\cellcolor{gray!40}\else
\ifdim#1pt<25pt\cellcolor{gray!50}\else
\ifdim#1pt<30pt\cellcolor{gray!60}\else
\ifdim#1pt<80pt\cellcolor{gray!70}\else
\ifdim#1pt<120pt\cellcolor{gray}\else
\cellcolor{white}\fi\fi\fi\fi\fi\fi\fi\fi\fi
#1}
\newcolumntype{C}{>{\collectcell\zz}c<{\endcollectcell}}

\ccsdesc[500]{Social and professional topics~Computer science education}
\ccsdesc[500]{Social and professional topics~Computing education programs}
\ccsdesc[500]{Social and professional topics~K-12 education}

\keywords{Programming education, primary school, teacher survey.}

\title[Challenging but Full of Opportunities]{Challenging but Full of Opportunities: \\Teachers' Perspectives on Programming in Primary Schools}

\author{Luisa Greifenstein}
\authornote{Both authors contributed equally to this research.}
\email{luisa.greifenstein@uni-passau.de}
\affiliation{%
  \institution{University of Passau}
  \state{Passau}
  \country{Germany}
}

\author{Isabella Graßl}
\authornotemark[1]
\email{isabella.grassl@uni-passau.de}
\affiliation{%
  \institution{University of Passau}
  \state{Passau}
  \country{Germany}
}

\author{Gordon Fraser}
\email{gordon.fraser@uni-passau.de}
\affiliation{%
  \institution{University of Passau}
  \city{Passau}
  \country{Germany}
}

\renewcommand{\shortauthors}{Greifenstein, Graßl \& Fraser}

\begin{abstract}
The widespread establishment of computational thinking in school curricula
requires teachers to introduce children to programming already at primary
school level. As this is a recent development, primary school teachers may
neither be adequately prepared for \emph{how} to best teach programming, nor
may they be fully aware \emph{why} they have to do so.
In order to gain a better understanding of these questions, we contrast
insights taken from practical experiences with the anticipations of teachers in
training. By surveying \numParticipantsPractice teachers who have taught
programming at primary schools and \numParticipantsTraining teachers in
training, we identify relevant challenges when teaching programming,
opportunities that arise when children learn programming, and strategies how to
address both of these in practice.
While many challenges and opportunities are correctly anticipated, we find
several disagreements that can inform revisions of the curricula in teaching
studies to better prepare primary school teachers for teaching programming at
primary schools.
\end{abstract}

%
%

\maketitle

\section{Introduction}
\label{sec:intro}


Programming is increasingly introduced at primary schools around the
world~\cite{heintz2016}. While secondary school teachers may be expected to be
adequately educated in their subjects and computing is usually a dedicated
subject, the education of primary school teachers who usually teach several
subjects without specialising on one particularly~\cite{rich2019coding} has
been reported to be insufficient~\cite{sentance2017computing}. In order to
support primary school teachers it is important to improve their education to
better prepare them for programming at primary schools, for example by teaching
them to program with
\Scratch~\cite{yukselturk2017investigation,fesakis2009influence,saez2020exploring}.

However, beyond acquiring basic programming skills, teacher education needs to
cover two further important aspects: First, primary school teachers need to be
adequately prepared for the challenges they may face in the classroom; in
particular, the challenges arising in practice may be different from those
anticipated while studying general education theory. Second, primary school
teachers also need to be taught why they are teaching programming to children
in the first place, so that they can appreciate the opportunities that may
arise for the children as a result of being able to program, and can take
measures in order to foster these.

We conducted a two phase survey in order to shed light on both of these
aspects. First, we surveyed \numParticipantsTraining teachers in training
asking them what challenges and opportunities they anticipate to encounter in
the classroom. Then, we surveyed \numParticipantsPractice teachers in practice
asking them which challenges and opportunities they actually encountered in
practice, and how they deal with the specific challenges and opportunities that
teachers in training most frequently anticipate. Overall, the concerns and
opportunities perceived by teachers in training represent a basis upon which
the content of teacher training can be improved. The differences in the
opinions of teachers in training and in practice can further be used to shape
the concept of teachers in training towards a more realistic view. Finally, the
ideas of experienced international teachers on how to deal with the challenges
and opportunities can enrich teacher training by explaining the suggested
strategies.

In detail, we aim to answer the following research questions using the two
surveys. First, a primary concern is what challenges teachers in practice
encountered in primary schools:

\smallskip
\noindent\textbf{RQ1: } \textit{What challenges have primary school teachers in practice encountered when teaching programming?}
\smallskip

\noindent By contrasting anticipation and reality, we want to inform primary
school teacher education about how to counter anticipated challenges as well as
how they align with practice. The second research question therefore is as
follows:

\smallskip
\noindent\textbf{RQ2: } \textit{What challenges do primary school teachers in training anticipate and what solutions do teachers in practice suggest?}
\smallskip
 
\noindent Effective teaching should also foster and reinforce opportunities
arising for children from learning to program. We therefore investigate the
following research question:

\smallskip
\noindent\textbf{RQ3: } \textit{What opportunities have primary school teachers in practice experienced when teaching programming?}
\smallskip

\noindent To see how current teacher education raises awareness of these opportunities,
and to improve teacher education by informing teachers in training how to
support the opportunities they anticipate, we ask the final research question:

\smallskip
\noindent\textbf{RQ4: } \textit{What opportunities do primary school teachers in training anticipate and what promotion do teachers in practice suggest?}
\smallskip

%

Our results show that many of the challenges and opportunities are correctly
anticipated by teachers in training, but there are several points of which they
are unaware, such as organisational issues at schools, or where they
overestimate importance, such as the complexity of programming.
These results can be used to improve teacher training with solutions for
existing concerns, central challenges, and possibilities to promote existing
opportunities.

\section{Related Work}
\label{sec:relatedwork}

\paragraph{Challenges of Teaching Programming}
Changes in the curriculum usually go along with issues for different stakeholders~\cite{ryder2015being}. This also applies to computer science education (CSE), as programming has often only recently been introduced at primary schools around the world~\cite{heintz2016}. Indeed, for some teachers it is difficult to include the mandatory computing program and therefore they rather neglect it~\cite{larke2019agentic}. This might be attributed to challenges that teachers face when teaching programming.
There are several studies that analyse challenges faced by CS teachers~\cite{sentance2017computing,yadav2016expanding,israel2015supporting,girvan2016extending,vinnervik2020implementing}. These focus on the respective national curricula and on different school levels, but some results seem to be shared: One common challenge is the lacking subject knowledge of teachers, which is mentioned even more often by primary school teachers than secondary school teachers~\cite{sentance2017computing}. This might be explained by primary school teachers covering several subjects, while secondary school teachers focus on fewer subjects in which they receive more in-depth training. 
%
%

\paragraph{Opportunities of Teaching Programming}
Few studies explicitly address the holistic potential of programming in primary school, and most highlight and analyse only partial aspects. A leading role is taken by the promotion of computational thinking (CT), which is considered to promote three prevalent framings: skill and competence building, creativity and social and ethical aspects~\cite{kafai2020theory}.
In particular, cognitive skills such as abstract and logical thinking or inhibition control are emphasised~\cite{cciftci2020effect, arfe2020effects}. These skills are especially important because students often demonstrate little algorithmic thinking~\cite{koulouri2014teaching, dagiene2014students}. 
 The potential of programming skills is often highlighted when considering future careers, as programming is not only important in pure software professions, but also in other industries \cite{vee2017coding}, and society in general~\cite{tuomi2018coding}.
Teachers in practice may not be aware of these scientific studies, therefore surveying teachers about their perceptions and experiences is important.
 
\paragraph{Effects of Teacher Training}
In order to prepare teachers to deal with changes in the CSE curriculum,
there are different approaches, mainly concerning training of software-based programming (mainly with \Scratch), followed by robotics, unplugged programming and game-based learning~\cite{ausiku2020preparing}.
Regardless of the technology used it is essential that teacher training provides experiences in practicing as well as teaching programming~\cite{mason2019preparing}. This can promote pedagogical content knowledge and self-efficacy and change pedagogical beliefs~\cite{mason2019preparing}. This is crucial as teachers in training might have obsolete prior experience from their time as students~\cite{ertmer2010teacher}. 
To be able to include the beliefs of teachers in training, they have to be identified first. While there is research on challenges perceived by teachers in practice and on the approaches used in teacher training, there is a lack of bridging these two aspects. In this paper, we combine the opinions of teachers in training and teachers in practice on challenges and opportunities when teaching programming. 
\section{Method}
\label{sec:methodology}


%
%

\subsection{Study Design}

In order to answer the research questions (cf. \cref{sec:intro}), we
conducted two surveys: One with teachers in training (\tt), and one with teachers in
practice (\tp), informed by the results of the former.
In both surveys, in addition to certain questions given below, all participants were asked about their demographic data, their previous programming experience and their attitudes towards teaching programming in primary school.

\subsubsection{Survey 1: Teachers in Training}

In the first survey, we asked primary school teachers \textit{in training} to
answer two open questions: 
\begin{asparaenum}
\item \emph{``What challenges do you see regarding teaching
programming in primary schools?''}
\item \emph{``What opportunities do you see regarding
programming in primary schools?''}
\end{asparaenum}

\subsubsection{Survey 2: Teachers in Practice}

In the second survey, we asked international primary school teachers \textit{in
practice} several questions. The first half of the questions dealt with
challenges, the second half with opportunities. The first question was the same
general question on perceived challenges as in the first survey:
\begin{asparaenum}
\item \emph{``What challenges do you see regarding teaching
programming in primary schools?''}
\end{asparaenum}
After that, we asked participants to rank the ten challenges that were most
frequently mentioned by the primary school \tt:
\begin{asparaenum}
\setcounter{enumi}{1}
\item \emph{``Please rank the following challenges regarding their relevance perceived by you.''}
\end{asparaenum}
Then, they were asked to pick the three challenges they perceive as most
relevant. For each of these three, they had to answer:
\begin{asparaenum}
\setcounter{enumi}{2}
\item  \emph{``How would you decrease
or master these challenges?''}
\end{asparaenum}
Finally, there is ongoing research on automated analysis tools to support teachers in addressing the challenges they encounter, but we do not anticipate that such tools are in widespread use yet. We therefore questioned the \tp whether they believe automated analysis tools can be useful in practice:
\begin{asparaenum}
\setcounter{enumi}{3}
\item 
\emph{``Automatic analysis tools can support teachers with giving feedback to their
students. (LitterBox, available at http://scratch-litterbox.org/, is an example
for an automatic code analysis tool which can find recurring bug patterns in
Scratch programs.)''} 
\end{asparaenum}
They had to state their agreement with this statement using a 5-point Likert scale and explain their rating.

The second half of the survey focused on
opportunities:
\begin{asparaenum}
\setcounter{enumi}{4}
\item \emph{``What opportunities do you see regarding
programming in primary schools?''}
\end{asparaenum}
The participants were asked to rank the ten opportunities that were most
frequently mentioned by the primary school \tt:
\begin{asparaenum}
\setcounter{enumi}{5}
\item \emph{``Please rank the following opportunities regarding their relevance perceived by you.''}
\end{asparaenum}
For the three opportunities they perceive as most relevant they then had
to answer:
\begin{asparaenum}
\setcounter{enumi}{6}
\item \emph{``How would you support or improve these opportunities?''}
\end{asparaenum}
A much discussed opportunity of programming education is the promotion of girls in computer science. In particular, a matter of debate is the question whether this opportunity can be supported with gender-homogeneous programming classes. Since we do not expect gender-homogeneous classes to be commonplace in primary school practice, we explicitly asked teachers about their opinion:
\begin{asparaenum}
\setcounter{enumi}{7}
\item 
\emph{``Gender-homogeneous programming classes help to improve the opportunity of
encouraging girls.'' }
\end{asparaenum}
Again they had to state their agreement on a 5-point Likert scale and explain their rating.

\subsection{Participants}

\subsubsection{Survey 1: Teachers in Training}

We implemented this survey into a course on primary mathematics teacher
education
at the University of Passau in January/February 2021. Participation
was voluntary and of \numParticipantsTrainingTotal \tt signed up in the course,
\numParticipantsTraining (\SI{91.8}{\percent} female, \SI{8.2}{\percent} male) between 18 and 45 years
(average 20.37) participated.

To assess the knowledge of the teachers, we asked them about their
self-experience in programming. Two-thirds of the \tt (\SI{67}{\percent}) stated that they
 had not taken any programming lessons themselves at school, while
the remaining third had (\SI{33}{\percent}). To learn about teachers' mental
attitudes regarding the introduction of first programming concepts in order to
identify any prior biases, we asked in a 5-point Likert scale whether
it is useful to introduce programming in primary school. Among the \tt, \SI{45.3}{\percent}
strongly agreed or somewhat agreed, while \SI{37.1}{\percent} were neutral.

\subsubsection{Survey 2: Teachers in Practice}

To elicit responses for the second survey, we used the Prolific platform\footnote{\url{https://prolific.co}}. As a
prescreening filter we restricted participants to teachers working in
primary/secondary (K-12) education, with a minimum approval rate of
90, which is the percentage of studies for which a participant has been approved. Using these filters, 1.377 of 147.942 individuals who
had been active in Prolific for the past 90 days matched our profile. In order
to find out if they had already taught programming in a primary school, we
designed a prescreening study with the single question of whether they had
already taught programming in primary school, and distributed it to these 1.377
people with a limit of 400 participants. Of these 400,
we were able to evaluate 397 responses, 251 of which answered our question in
the affirmative and were thus eligible for the main study. The compensation was based on an average wage of £7.50 per hour.
The main study was sent to the 251 people from the prescreening study in
June/July 2021, achieving a total of 209 respondents. Participants were compensated with £2.50 (based on an average hourly wage of £10.54). We excluded 9 participants because they did not answer all questions.

The \numParticipantsPractice (\SI{79}{\percent} female, \SI{21}{\percent} male) experienced
international primary school \tp were between 20 and 63
years of age (average 36.54). 
A large majority of
the \tp currently reside in the UK (\SI{70}{\percent}), \SI{17}{\percent} in the rest of Europe, and \SI{13}{\percent} in the rest of the world.

To better assess the knowledge of the teachers, we asked them about their self-experience in programming and their previous teaching in primary schools.
Around half of the \tp (\SI{53}{\percent}) stated that they do not have any prior programming experience. 
Those who program themselves do this in \Scratch~\cite{resnick2009scratch} (\SI{25.5}{\percent}), Java (\SI{21.5}{\percent}) or Python (\SI{17.5}{\percent}). In the
classroom, most \tp use \Scratch (\SI{65.5}{\percent}). \Purplemash\footnote{\url{https://www.purplemash.com}}
is also popular with a
third (\SI{31.5}{\percent}) and exactly a quarter use \textit{unplugged} programming~\cite{bell2015cs}. \SI{19.5}{\percent} of the \tp also use \Codeorg. Nearly half (\SI{43.5}{\percent}) of \tp teach programming
for 1 hour in primary grades, and \SI{27.5}{\percent} teach it for only 30 minutes per week.
For most, programming classes are compulsory (\SI{70.5}{\percent}). Compared to the
\tt, the \tp are more strongly in favour of introducing first programming concepts: \SI{91}{\percent} agreed or strongly agreed that it is useful to teach
programming and \SI{4}{\percent} were neutral.

\subsection{Data Analysis} 
The open questions provided us with qualitative data on which we applied
thematic analysis~\cite{bergman2010hermeneutic}: For each question, we first
collected themes, then counted them and in a final step again related them to
the original data and our research questions. To ensure inter-rater reliability
two raters classified the first 20 statements regarding each open question and
agreed on a coding scheme. Each rater then classified half of all statements.
Additionally, 40 statements per question were rated by both raters to measure the inter-rater agreement, which is good
 at $K$ = \irr.\footnote{For replications, all data of the study including coding schemes are available at \url{https://github.com/se2p/study-teacher-chall-opps}.}
To answer RQ1 (challenges) and RQ3 (opportunities) we consider the percentage of \tp who mentioned the respective code at least once.
To answer RQ2 and RQ4 we consider three aspects: First, we compare the percentage of \tt who mentioned the respective code with the percentage of \tp. We measure statistical differences using a Wilcoxon Rank Sum test with $\alpha = 0.05$. Second, we rank the ten challenges/opportunities mentioned most frequently by the \tt based on the percentage of the \tp who included them in the three that they deemed important and provided a response for. Third, for the question on analysis tools and gender homogeneous teaching we consider both, the Likert-scale data and the codes resulting from the responses.

\subsection{Threats to Validity}

\noindent\emph{External validity:} 
Survey 1 is based on a self-selecting sample as participation was not required to pass the university course, and all \tt were from the same university. 
For survey 2, a large proportion of the \tp are from the UK, which is likely
because programming has not yet been introduced in primary schools in many
countries and thus experiences may be lacking. However, survey 2
nevertheless represents a broad international spectrum.

\noindent\emph{Internal validity:} 
The \tt were given a shorter questionnaire to complete in their spare time,
while the \tp were monetarily compensated for the more extensive questionnaire. 
Two authors independently annotated the data,
developed categories, reviewed the coding scheme and discussed the few
non-matches in detail.

\noindent\emph{Construct validity:} 
As the survey relies on the authenticity of the respondents,
the subjective impressions and experiences are rather to been seen as a guideline. Several researchers besides the authors independently reviewed the questionnaire to reduce the risk of misinterpreting the questions.

\section{Results}

\begin{table}[t]\centering
	\caption{Perceived challenges by \tp and \ttEnd. \\
	\textmd{\small The percentage corresponds to the proportion of \tp and \tt mentioning the subcategory at least once in the open question.}}
	\vspace{-1em}
	\label{tab:challengesCodingScheme}
	\setlength{\tabcolsep}{0.2em}
	\resizebox{\columnwidth}{!}{

\begin{tabular}{lp{4.5cm}rlrl}
	\toprule
	(Sub-) \\ Category & Themes & \% TP & \hspace{\MAX} & \% TT & \hspace{\MAX} \\
		 \midrule
	\rowcolor{Gray}
	
		\multicolumn{2}{l}{School}  & \printpercent{\PercentPracticeChallengesSchool} & \DrawPercentageBar{\PercentPracticeChallengesSchool} & \printpercent{\PercentTeacherChallengesSchool} & \DrawPercentageBar{\PercentTeacherChallengesSchool} \\
		 organisational & media equipment, funding, internet connection  & \printpercent{\PercentPracticeChallengesSchoolorganisational} & \DrawPercentageBar{\PercentPracticeChallengesSchoolorganisational} & \printpercent{\PercentTeacherChallengesSchoolorganisational} & \DrawPercentageBar{\PercentTeacherChallengesSchoolorganisational} \\
	
	\rowcolor{Gray}
	\multicolumn{2}{l}{Students} & \printpercent{\PercentPracticeChallengesStudents} & \DrawPercentageBar{\PercentPracticeChallengesStudents} & \printpercent{\PercentTeacherChallengesStudents} & \DrawPercentageBar{\PercentTeacherChallengesStudents} \\
	cognitive & prior media experience, overwhelming, reasoning, literacy, subject knowledge  & \printpercent{\PercentPracticeChallengesStudentscognitive} & \DrawPercentageBar{\PercentPracticeChallengesStudentscognitive} & \printpercent{\PercentTeacherChallengesStudentscognitive} & \DrawPercentageBar{\PercentTeacherChallengesStudentscognitive} \\
	affective & interest and motivation  & \printpercent{\PercentPracticeChallengesStudentsaffective} & \DrawPercentageBar{\PercentPracticeChallengesStudentsaffective} & \printpercent{\PercentTeacherChallengesStudentsaffective} & \DrawPercentageBar{\PercentTeacherChallengesStudentsaffective} \\
	heterogeneity & prior media experience, interest, subject knowledge, gender  & \printpercent{\PercentPracticeChallengesStudentsheterogeneity} & \DrawPercentageBar{\PercentPracticeChallengesStudentsheterogeneity} & \printpercent{\PercentTeacherChallengesStudentsheterogeneity} & \DrawPercentageBar{\PercentTeacherChallengesStudentsheterogeneity} \\
	metacognitive & distractability, concentration, impatience & \printpercent{\PercentPracticeChallengesStudentsmetacognitive} & \DrawPercentageBar{\PercentPracticeChallengesStudentsmetacognitive} & \printpercent{\PercentTeacherChallengesStudentsmetacognitive} & \DrawPercentageBar{\PercentTeacherChallengesStudentsmetacognitive} \\
	young age  & & \printpercent{\PercentPracticeChallengesStudentsyoungage} & \DrawPercentageBar{\PercentPracticeChallengesStudentsyoungage} & \printpercent{\PercentTeacherChallengesStudentsyoungage} & \DrawPercentageBar{\PercentTeacherChallengesStudentsyoungage} \\
	
	\rowcolor{Gray}
	\multicolumn{2}{l}{Teachers} & \printpercent{\PercentPracticeChallengesTeachers} & \DrawPercentageBar{\PercentPracticeChallengesTeachers} & \printpercent{\PercentTeacherChallengesTeachers} & \DrawPercentageBar{\PercentTeacherChallengesTeachers} \\
	cognitive & subject knowledge, overwhelming, prior media experience & \printpercent{\PercentPracticeChallengesTeacherscognitive} & \DrawPercentageBar{\PercentPracticeChallengesTeacherscognitive} & \printpercent{\PercentTeacherChallengesTeacherscognitive} & \DrawPercentageBar{\PercentTeacherChallengesTeacherscognitive} \\
	didactic & individual support, child-friendly implementation, prevention of distraction & \printpercent{\PercentPracticeChallengesTeachersmethodicalanddidactic} & \DrawPercentageBar{\PercentPracticeChallengesTeachersmethodicalanddidactic} & \printpercent{\PercentTeacherChallengesTeachersmethodicalanddidactic} & \DrawPercentageBar{\PercentTeacherChallengesTeachersmethodicalanddidactic} \\
	affective & self-efficacy & \printpercent{\PercentPracticeChallengesTeachersaffective} & \DrawPercentageBar{\PercentPracticeChallengesTeachersaffective} & \printpercent{\PercentTeacherChallengesTeachersaffective} & \DrawPercentageBar	{\PercentTeacherChallengesTeachersaffective} \\
	
	\rowcolor{Gray}
	\multicolumn{2}{l}{Programming} & \printpercent{\PercentPracticeChallengesProgramming} & \DrawPercentageBar{\PercentPracticeChallengesProgramming} & \printpercent{\PercentTeacherChallengesProgramming} & \DrawPercentageBar{\PercentTeacherChallengesProgramming} \\
	complexity & programming language, technical terms, abstract, relation to life & \printpercent{\PercentPracticeChallengesProgrammingcomplexity} & \DrawPercentageBar{\PercentPracticeChallengesProgrammingcomplexity} & \printpercent{\PercentTeacherChallengesProgrammingcomplexity} & \DrawPercentageBar{\PercentTeacherChallengesProgrammingcomplexity} \\
	problems & debugging & \printpercent{\PercentPracticeChallengesProgrammingdebuggingandproblems} & \DrawPercentageBar{\PercentPracticeChallengesProgrammingdebuggingandproblems} & \printpercent{\PercentTeacherChallengesProgrammingdebuggingandproblems} & \DrawPercentageBar{\PercentTeacherChallengesProgrammingdebuggingandproblems} \\
	
	\rowcolor{Gray}
	\multicolumn{2}{l}{Government} & \printpercent{\PercentPracticeChallengesGovernment} & \DrawPercentageBar{\PercentPracticeChallengesGovernment} & \printpercent{\PercentTeacherChallengesGovernment} & \DrawPercentageBar{\PercentTeacherChallengesGovernment} \\
	organisational & time, curriculum & \printpercent{\PercentPracticeChallengesGovernmentorganisational} & \DrawPercentageBar{\PercentPracticeChallengesGovernmentorganisational} & \printpercent{\PercentTeacherChallengesGovernmentorganisational} & \DrawPercentageBar{\PercentTeacherChallengesGovernmentorganisational}  \\
	
	\rowcolor{Gray}
	\multicolumn{2}{l}{Parents} &  \printpercent{\PercentPracticeChallengesParents}	& \DrawPercentageBar{\PercentPracticeChallengesParents} &  \printpercent{\PercentTeacherChallengesParents}	& \DrawPercentageBar{\PercentTeacherChallengesParents} \\
	organisational & media equipment  & \printpercent{\PercentPracticeChallengesParentsorganisational}	& \DrawPercentageBar{\PercentPracticeChallengesParentsorganisational} & \printpercent{\PercentTeacherChallengesParentsorganisational}	& \DrawPercentageBar{\PercentTeacherChallengesParentsorganisational} \\
	affective & fear and criticism & \printpercent{\PercentPracticeChallengesParentsaffective} & \DrawPercentageBar{\PercentPracticeChallengesParentsaffective} & \printpercent{\PercentTeacherChallengesParentsaffective} & \DrawPercentageBar{\PercentTeacherChallengesParentsaffective} \\

\bottomrule
\end{tabular}
}
\end{table}

\subsection{RQ1: Challenges Experienced by \tp}
\label{sec:rq1}

To answer RQ1 we consider the responses to the open question on challenges experienced by the \tp (survey 2).
\Cref{tab:challengesCodingScheme} shows the identified
categories with their subcategories and themes.

\subsubsection{School}
The most frequently named challenge with \printpercent{\PercentPracticeChallengesSchool} of \tp is concerned with schools being organisationally
challenged. In particular,
this can be split into three individual challenges captured in this
response:
\tpquote{189}{school funding issues, particularly around access to resources or even ensuring a consistent internet connection.}

\subsubsection{Students}
The second most frequently mentioned category with \printpercent{\PercentPracticeChallengesStudents} of \tp relates to challenges that primary school
students might face when programming. 
Cognitive issues are mentioned most often and partially explained by a lack of digital literacy: 
\tpquote{19}{With the younger students it is difficult for them to learn and implement many instructions because they are often still learning basic computer skills like how to use a mouse [...].}
The use of digital media is also linked to metacognitive challenges: 
\tpquote{173}{Children are easily distracted when given an exciting new tool to play with, so keeping them on task is a challenge in itself.}
Affective factors of students can be even in
conflict with other challenges such as 
\tpquote{2}{Keeping the programming simple, but still having interesting outcomes to keep the students engaged.}
Moreover, all the mentioned characteristics differ between students which leads to the challenge of
\tpquote{76}{trying to teach the whole class at the same time with so many different abilities.}
Even if the students might not be heterogeneous in terms of age,
\tpquote{74}{It's very difficult for younger children to grasp.}

\subsubsection{Teachers}

Challenges related to teachers are named by \printpercent{\PercentPracticeChallengesTeachers} of \tp. In the context of didactic issues, the most often mentioned challenge regards the individual support: 
\tpquote{75}{younger children tend to need a lot of help with this and sometimes it's hard to get round them all.}
Studies on debugging confirm that also high-school teachers struggle with the rush when helping students~\cite{michaeli2021debugging}. Such situations might
in turn exacerbate the challenge of low self-efficacy. However, cognitive issues are perceived as the greatest challenge (\cref{tab:challengesCodingScheme}) of
teachers such as 
\tpquote{23}{limited and outdated subject knowledge within
teaching staff.}
This matches prior research indicating lack of subject knowledge as a
major
challenge~\cite{sentance2017computing,yadav2016expanding,
vinnervik2020implementing}.

\subsubsection{Programming}
A total of \printpercent{\PercentPracticeChallengesProgramming} of \tp name challenges related to programming itself: 
\tpquote{123}{At first they have some difficulties understanding the concept, especially that every single step needs to be detailed (as opposed to when you explain something to a human [...])
}
%
and \tpquote{12}{The debugging aspect, which is actually a very powerful learning experience, causes lots of initial impatience!}
Thus, programming is related to both cognitive and affective challenges.

\subsubsection{Government}
A total of \printpercent{\PercentPracticeChallengesGovernment} of \tp see organisational issues regarding time and curricula, and some participants even explicitly connected these related themes: 
\tpquote{41}{time allowable in the curriculum for children to understand programming as it's not an easy task for all to accomplish satisfactorily in the time given.}

\subsubsection{Parents}
They are hardly mentioned (\printpercent{\PercentPracticeChallengesParents}
of \tp), and mainly regarding media equipment:
\tpquote{114}{Some children in disadvantaged schools dont [sic] have easy access to computer systems at home.}

\summary{RQ1}{The main challenges of \tp are organisational issues of the school, and cognitive issues of teachers and students.
}


\subsection{RQ 2: Challenges Perceived by \tt and Strategies for these from \tp}
\label{sec:rq2}

\begin{table}[t]\centering
	\caption{Top ten challenges of \tt ranked by \tpEnd.	\\
	\textmd{The \% \tp corresponds to the proportion of \tp ranking the respective challenge among the three most relevant challenges.}}
	\vspace{-1em}
	\label{tab:challsTen}
	\setlength{\tabcolsep}{0.2em}
	\resizebox{\columnwidth}{!}{
\begin{tabular}{rlp{6cm}rl}
	\toprule
	Nr. & Category & Challenge 
	& \% TP & \\
	\midrule

	1 & School & There might be a lack of technical equipment at primary schools. 
& \printpercent{\ChallengePercentSix} & \DrawPercentageBar{\ChallengePercentSix}
	\\
	
	2 & Teachers & Teachers might be challenged with cognitive issues (e.g. because of a lack of subject knowledge). 
	& \printpercent{\ChallengePercentThree} & \DrawPercentageBar{\ChallengePercentThree}
	\\

	3 & Government & Institutions might be organizationally challenged (e.g. because of an overloaded curriculum and programming being time consuming). 
& \printpercent{\ChallengePercentTen} & \DrawPercentageBar{\ChallengePercentTen}
	\\

	4 & Teachers & Teachers might not know which didactic considerations and methods they should use to teach programming (e.g. how to master looking after each student and their individual programming issues). 
& \printpercent{\ChallengePercentFour} & \DrawPercentageBar{\ChallengePercentFour}
	\\

	5 & Teachers & Teachers might have negative attitudes (e.g. because of their self-concept towards programming). 
& \printpercent{\ChallengePercentTwo} & \DrawPercentageBar{\ChallengePercentTwo}
	\\
	
	6 & Programming & Programming might be a (too) complex topic for primary schools. 
& \printpercent{\ChallengePercentFive} & \DrawPercentageBar{\ChallengePercentFive}
	\\

	7 & Students & Students might be cognitively overwhelmed and might have insufficient prior knowledge regarding digital literacy. 
& \printpercent{\ChallengePercentEight} & \DrawPercentageBar{\ChallengePercentEight}
	\\
	
	8 & Students& Students might have problems because of metacognitive issues (e.g. by being easily distracted). 
	& \printpercent{\ChallengePercentNine} & \DrawPercentageBar{\ChallengePercentNine}
	\\

	9 & Parents & Parents might have negative attitudes or there might be a lack of technical equipment at home. 
& \printpercent{\ChallengePercentOne} & \DrawPercentageBar{\ChallengePercentOne}
	\\

	10 & Students & There might be differences between the students (resulting e.g. from girls having different tendencies or preferences than boys). 
& \printpercent{\ChallengePercentSeven} & \DrawPercentageBar{\ChallengePercentSeven}
	\\

\bottomrule
\end{tabular}
}
\end{table}

\begin{table}[t]
\centering
\caption{Strategies of TP for the challenges of TT. \\
\textmd{The absolute values refer to all mentions and the percentage to the proportion of TP mentioning the subcategory at least once.}} 
	\vspace{-1em}
	\setlength{\tabcolsep}{0.2em}
	\label{tab:challheatmap}
	\resizebox{\columnwidth}{!}{
\begin{tabular}{llCCCCCCCCCCCr}
\toprule
Category                                       & Subcategory                  & \multicolumn{1}{c}{1} & \multicolumn{1}{c}{2} & \multicolumn{1}{c}{3} & \multicolumn{1}{c}{4} & \multicolumn{1}{c}{5} & \multicolumn{1}{c}{6} & \multicolumn{1}{c}{7} & \multicolumn{1}{c}{8} & \multicolumn{1}{c}{9} & \multicolumn{1}{c}{10} & \multicolumn{1}{c}{$\sum$ } & \% \tp \\
\midrule
Teachers                    & participate in training      & 0   & 101  & 1  & 45   & 0    & 29   & 1  & 3  & 0  & 0 & 180 &  \printpercent{\StrategyPercentStrategiesforChallengesTeachersparticipateintraining}              \\
                            & teaching & 12  & 2    & 10 & 11   & 0    & 1    & 29 & 26 & 21 & 5 & 117 & \printpercent{\StrategyPercentStrategiesforChallengesTeachersteaching}               \\
                            & content of training          & 0   & 26   & 1  & 14   & 0    & 23   & 0  & 0  & 0  & 0 & 64 & \printpercent{\StrategyPercentStrategiesforChallengesTeacherscontentoftraining}               \\
                            & cooperate with others        & 4   & 6    & 2  & 7    & 0    & 6    & 3  & 3  & 0  & 0 & 31 & \printpercent{\StrategyPercentStrategiesforChallengesTeacherscooperatewithothers}                \\
                            & schedule time                         & 0   & 0    & 11 & 1    & 0    & 0    & 9  & 4  & 2  & 1 & 28 & \printpercent{\StrategyPercentStrategiesforChallengesTeacherstime}                \\
                            & use existing material & 0   & 9    & 3  & 5    & 0    & 4    & 0  & 0  & 0  & 0 & 21 & \printpercent{\StrategyPercentStrategiesforChallengesTeachersgetsupportthroughmaterial}                \\

School                      & media equipment              & 96  & 0    & 0  & 2    & 4    & 0    & 0  & 1  & 0  & 0 & 103 & \printpercent{\StrategyPercentStrategiesforChallengesSchoolmediaequipment}              \\                       
                            & course after school          & 0   & 0    & 7  & 0    & 2    & 0    & 0  & 0  & 0  & 0 & 9 & \printpercent{\StrategyPercentStrategiesforChallengesSchoolcourseafterschool}             \\
                            & joint approach               & 0   & 0    & 5  & 2    & 0    & 0    & 1  & 0  & 0  & 0 & 8 & \printpercent{\StrategyPercentStrategiesforChallengesSchooljointapproach}                 \\
                            & responsible teacher          & 1   & 2    & 3  & 0    & 0    & 0    & 0  & 0  & 0  & 0 & 6 & \printpercent{\StrategyPercentStrategiesforChallengesSchoolresponsibleteacher}                 \\

Government                  & more funds for CSE           & 51  & 0    & 2  & 0    & 1    & 0    & 0  & 0  & 0  & 0 & 54 & \printpercent{\StrategyPercentStrategiesforChallengesGovernmentmorefundsforCSE}               \\
                            & reorganise curricula         & 0   & 0    & 29 & 1    & 0    & 0    & 7  & 11 & 0  & 0 & 48 & \printpercent{\StrategyPercentStrategiesforChallengesGovernmentreorganisecurricula}               \\

Parents                    & content                      & 0   & 0    & 0  & 0    & 10   & 0    & 3  & 0  & 0  & 0 & 13 & \printpercent{\StrategyPercentStrategiesforChallengesParentsContent}                 \\
                            & meetings                     & 0   & 1    & 1  & 0    & 6    & 0    & 1  & 0  & 0  & 0 & 9 & \printpercent{\StrategyPercentStrategiesforChallengesParentsMeetings}            \\
                            & support                      & 0   & 0    & 0  & 0    & 6    & 0    & 0  & 0  & 0  & 0 & 6 & \printpercent{\StrategyPercentStrategiesforChallengesParentsSupport}                        \\

Society                     & rethink sociologically       & 0   & 0    & 0  & 0    & 0    & 0    & 0  & 0  & 0  & 2 & 4 & \printpercent{\StrategyPercentStrategiesforChallengesSocietyrethinksociologically}                \\

$\sum$            &                              & 164 & 147  & 75 & 88   & 29   & 63   & 54 & 48 & 23 & 8 & 1400   \\

\bottomrule
\end{tabular}
}
\end{table}

\begin{figure}[tb]
	\centering
	\includegraphics[width=\columnwidth]{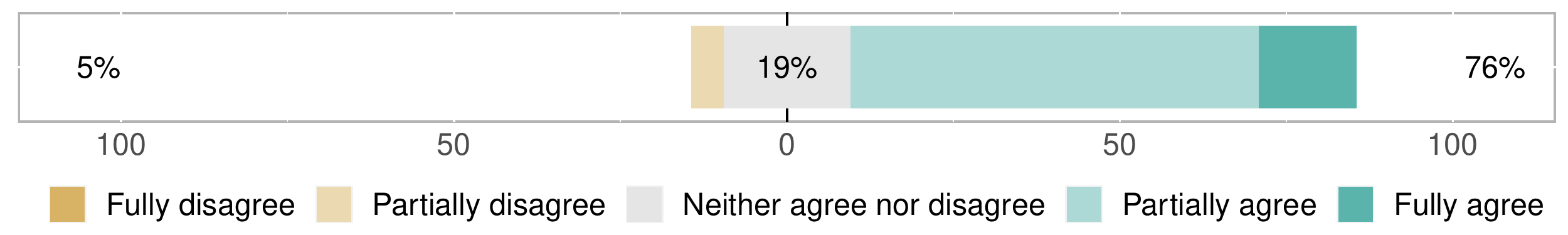}
	\vspace{-2em}
	\caption{\label{fig:chall_analysistools_likert}Opinions of \tp on analysis tools.}
\end{figure}

To answer RQ2 we consider the open question on the perceived challenges (survey 1 and 2), as well as the remaining questions on challenges and strategies of survey 2.
\cref{tab:challsTen} shows the ten most common challenges mentioned by \tt, ranked by \tp.
\cref{tab:challheatmap} includes the strategies that \tp mentioned for the top ten challenges of \tt.
\cref{fig:chall_analysistools_likert} shows the rating of analysis tools. 

\subsubsection{Challenges Perceived by \ttEnd}
All categories and subcategories are mentioned by both \tt and \tp (\cref{tab:challengesCodingScheme}). However, the distribution differs, which can be explained by significant differences in the subcategories. In the following we discuss these differences.

\paragraph{Students}
\tt mentioned cognitive challenges significantly more often (\pvalue{\PValueChallengesStudentscognitive}). This indicates that students are less cognitively overwhelmed and have more sufficient media experience and subject knowledge than \tt assume. Interestingly, the theme ``reasoning'' is mentioned only by \tp which is a more general cognitive skill and associated with CT \cite{selby2014refining}. 
A similar picture emerges regarding the challenge of heterogeneity: \tt consider this aspect significantly more often as challenging than \tp (\pvalue{\PValueChallengesStudentsheterogeneity}). This might also be attributed to only \tt mentioning gender differences as an issue.

\paragraph{Teachers}
\tt mentioned challenges regarding the teachers slightly more often (\cref{tab:challengesCodingScheme}). \tt perceive the didactic issues as significantly more challenging than \tp (\pvalue{\PValueChallengesTeachersmethodicalanddidactic}). Consequently, \tp do not seem to be overwhelmed to choose an appropriate teaching approach. Therefore, \tt can benefit from the strategies and working methods used by \tp that are explained in \cref{sec:rq2,sec:rq4}.

\paragraph{Programming}
\tt 
put a different focus (\cref{tab:challengesCodingScheme}): \tt mentioned the complexity of programming significantly more often (\pvalue{\PValueChallengesProgrammingcomplexity}) and \tp the debugging process when teaching programming (\pvalue{\PValueChallengesProgrammingdebuggingandproblems}). This is why debugging should be supported in the classroom~\cite{michaeli2021debugging} or within teacher training~\cite{greifenstein2021effects}.

\paragraph{School}
\tt seem to underestimate the organisational issues of schools as \tp mentioned them significantly more often (\pvalue{\PValueChallengesSchoolorganisational}). This can be attributed to \tt not having considered funding and internet connection but only the issue of media equipment.

\paragraph{Government}
Regarding the government, there are no significant differences: Both \tt and \tp mentioned curriculum and time issues.

\paragraph{Parents}
Although parents are mentioned least often by both groups, \tt considered them almost three times as often (with \printpercent{\PercentTeacherChallengesParents} vs. \printpercent{\PercentPracticeChallengesParents}) and particularly overestimated affective issues (\pvalue{\PValueChallengesParentsaffective}).

\subsubsection{Strategies}

The ten most common challenges mentioned by \tt were ranked regarding their relevance by the \tp (\cref{tab:challsTen}). 
For the three challenges that the \tp ranked most relevant, they were asked to describe a strategy (\cref{tab:challheatmap}). In the following, we explain for each challenge what strategies are suggested.

\paragraph{Challenge 1: Lack of equipment at schools}
For the most relevant challenge (\cref{tab:challsTen}) both school and governmental strategies are considered: The government should provide more funds for CSE  
and furthermore, \tp suggest different possibilities to get media equipment beyond simply buying it such as:
\tpquote{136}{Schools should also hold fund raising events---the children will often be happy to take part in events that will raise money for technology or `toys' for them to use.}
%
%
A different approach considers dividing students into groups which leads to a reduced need of technical equipment.

\paragraph{Challenge 2: Cognitive issues of teachers}
In order to counteract the lacking subject knowledge, \tp suggest participating in training that addresses content knowledge but also affective factors. Indeed, training with 68 \% of \tp is the most often mentioned strategy across all challenges (\cref{tab:challheatmap}). According to \tp, training should be more often, high-quality, for free, mandatory and prior to teaching.

\paragraph{Challenge 3: Organisationally challenged institutions}
The lack of time implies that curricula should be reorganised by 
\tpquote{26}{Introducin [sic] digital thinking in earlier years}
and
\tpquote{147}{Decrease or make other priorities less time consuming.} 
Further ideas are listed in \cref{tab:challheatmap}. In this context, the strategy of \didacticconsiderations mostly deals with teaching interdisciplinary as 
\tpquote{201}{Coding could be linked to Literacy/Numeracy learning in some way---that would benefit motivation in those subjects and offer time to do it.}

\paragraph{Challenge 4: Didactic issues of teachers} 
\tp suggest to participate in teacher training which focuses on pedagogical content knowledge rather than content knowledge. Besides training, within \didacticconsiderations simply trying out different methods and \tpquote{12}{peer mentoring among teaching staff} are suggested.

\paragraph{Challenge 5: Affective issues of teachers}
Interestingly, as for cognitive issues, teacher training should both address content knowledge and affective factors such as attitudes, motivation and confidence. This might be explained by well-informed teachers being confident:
\tpquote{38}{[...] an expert came to talk to the staff about `Scratch' and showed us how to do a variety of things. This certainly helped boost the confidence of the less technology literate members of staff.}

\paragraph{Challenge 6: Complexity of programming}
This might be reduced by \didacticconsiderations such as 
\tpquote{160}{Making sure we can simplify the concepts to relate to real life examples first, before going heavy on the programming concepts.}
This matches the strategy of ``contextualisation'' found by Sentance and Csizmadia~\cite{sentance2017computing}.

\paragraph{Challenge 7: Cognitive issues of students}
All student-related challenges (7, 8 and 10) are expected to be improved by the \didacticconsiderations of considering individual factors such as interest or strengths, e.g., by differentiating learning material. Cognitive issues in particular are often stated to decrease when the students get the chance of expanding their prior knowledge: 
\tpquote{148}{We would not introduce them straight to programming; instead, teach them how to better use technical equipment first.} 
Programming with hardware is addressed here, the explained previous promotion of digital literacy however is not relevant for programming unplugged \cite{bell2015cs}.

\paragraph{Challenge 8: Metacognitive issues of students}
Besides considering individual factors, metacognitive issues can be counteracted with increasing the students' engagement by, e.g., a
\tpquote{160}{distract free environment.}
and
\tpquote{198}{interactive ativities [sic].}

\paragraph{Challenge 9: Affective and organisational issues of parents}
Comprehensibly, this challenge is connected to all strategies regarding parents (\cref{tab:challheatmap}). \tp explained that 
\tpquote{15}{for the negative attitudes of parents, they can be invited for a day to see how their children work and how much fun they have}
or to
\tpquote{139}{create easy to follow guides to allow the parents to acquire knowledge.}

\paragraph{Challenge 10: Heterogeneity of students}
Besides considering individual factors when teaching, there might be a need for rethinking sociological views:
\tpquote{4}{Girs [sic] are usually raised for household chores and not more mental issues. this is a societal problem.}

\subsubsection{Automated Analysis Tools}

There exist several tools that analyse block-based programs~\cite{alves2019approaches}. Recent approaches for \Scratch programs deal with, e.g., next-step hints~\cite{obermuller2021guiding} or code perfumes~\cite{obermuller2021code}.
However, \tp do not
appear to be well aware of the potential of automated tools to support
programming education, as
\printpercent{\StrategyPercentAnalysisToolsuncertaintynoexperience} of \tp
explicitly stated that they had no experience with such tools.

However, \cref{fig:chall_analysistools_likert} shows that \tp like the idea and
consider automated tools potentially useful in terms of giving feedback. Giving
feedback can be located in challenge 4 (didactic issues of teachers) as individual
support is mentioned most often here  (\cref{tab:challengesCodingScheme}).
Besides giving and perceiving feedback, debugging, diagnosis, support and assessment are mentioned. This goes along with several explained advantages such as dealing with many students at the same time, feedback being immediate or more confidence of teachers. Consequently, these advantages might also help with cognitive and affective issues of teachers and the often described lack of time~\cite{vinnervik2020implementing,sentance2017computing}. 
\tpquote{170}{With so many students, and so much to do in the classroom, these tools allow teachers to provide feedbac [sic] to their students without needing to spend several hours reading code.}
Indeed, automatically generated hints can support teachers with debugging programs if the hints fulfil certain criteria~\cite{greifenstein2021effects}.

Some \tp reveal scepticism against tools and believe one should rather rely on human feedback:
\tpquote{127}{I don't think it can give a subjective feedback.} 
However, the lack of individualisation implies one advantage of
computer-based feedback: Tools give objective feedback---independent of, e.g., gender
or social background \cite{hattie2008visible}.

\summary{RQ2}{\tt partially consider other challenges than \tp, e.g., regarding the difficulty of programming. Generally, \tp regard teacher training and \didacticconsiderations as most helpful. Analysis tools are considered useful for giving efficient feedback.}

\subsection{RQ3: Opportunities Experienced by \tp}
\label{sec:rq3}

\begin{table}[t]\centering
	\caption{Perceived opportunities of \tp and \tt.\\
	\textmd{\small The percentage corresponds to the proportion of \tp and \tt mentioning the subcategory at least once in the open question.}
	} 
	\vspace{-1em}
	\setlength{\tabcolsep}{0.2em}
	\label{tab:opportunitiesCodingScheme}
	\resizebox{\columnwidth}{!}{
\begin{tabular}{lp{3.5cm}rlrl}	
\toprule
	(Sub-) \\ Category  & Themes & \% TP & \hspace{\MAX} & \% TT &  \\
\midrule

\rowcolor{Gray}
Skills acquisition &  & \printpercent{\PercentPracticeOpportunitiesSkillsacquisition} & \DrawPercentageBar{\PercentPracticeOpportunitiesSkillsacquisition} & \printpercent{\PercentTeacherOpportunitiesSkillsacquisition} & \DrawPercentageBar{\PercentTeacherOpportunitiesSkillsacquisition} \\
 cognitive & problem solving and logical thinking, linguistic stimulation &   \printpercent{\PercentPracticeOpportunitiesSkillsacquisitioncognitive} & \DrawPercentageBar{\PercentPracticeOpportunitiesSkillsacquisitioncognitive} & \printpercent{\PercentTeacherOpportunitiesSkillsacquisitioncognitive} & \DrawPercentageBar{\PercentTeacherOpportunitiesSkillsacquisitioncognitive} \\

holistic & creativity, digital literacy, computational literacy &  \printpercent{\PercentPracticeOpportunitiesSkillsacquisitionholistic} & \DrawPercentageBar{\PercentPracticeOpportunitiesSkillsacquisitionholistic} & \printpercent{\PercentTeacherOpportunitiesSkillsacquisitionholistic} & \DrawPercentageBar{\PercentTeacherOpportunitiesSkillsacquisitionholistic}  \\

affective & interest and motivation, reduction of prejudices, fun, self-confidence, talent &  \printpercent{\PercentPracticeOpportunitiesSkillsacquisitionaffective} & \DrawPercentageBar{\PercentPracticeOpportunitiesSkillsacquisitionaffective} & \printpercent{\PercentTeacherOpportunitiesSkillsacquisitionaffective} & \DrawPercentageBar{\PercentTeacherOpportunitiesSkillsacquisitionaffective}  \\
 
metacognitive & self-reliance &  \printpercent{\PercentPracticeOpportunitiesSkillsacquisitionmetacognitiveselfreliance} & \DrawPercentageBar{\PercentPracticeOpportunitiesSkillsacquisitionmetacognitiveselfreliance} & \printpercent{\PercentTeacherOpportunitiesSkillsacquisitionmetacognitiveselfreliance} & \DrawPercentageBar{\PercentTeacherOpportunitiesSkillsacquisitionmetacognitiveselfreliance}  \\
 
  \rowcolor{Gray}
Foundation &  & \printpercent{\PercentPracticeOpportunitiesTransferableknowledge} & \DrawPercentageBar{\PercentPracticeOpportunitiesTransferableknowledge}& \printpercent{\PercentTeacherOpportunitiesTransferableknowledge} & \DrawPercentageBar{\PercentTeacherOpportunitiesTransferableknowledge}\\
professional life &  &  \printpercent{\PercentPracticeOpportunitiesTransferableknowledgeprofessionallife} & \DrawPercentageBar{\PercentPracticeOpportunitiesTransferableknowledgeprofessionallife} &   \printpercent{\PercentTeacherOpportunitiesTransferableknowledgeprofessionallife} & \DrawPercentageBar{\PercentTeacherOpportunitiesTransferableknowledgeprofessionallife}  \\
 general future &  &  \printpercent{\PercentPracticeOpportunitiesTransferableknowledgegeneralfuture} & \DrawPercentageBar{\PercentPracticeOpportunitiesTransferableknowledgegeneralfuture} &   \printpercent{\PercentTeacherOpportunitiesTransferableknowledgegeneralfuture} & \DrawPercentageBar{\PercentTeacherOpportunitiesTransferableknowledgegeneralfuture}  \\
 secondary school &  &  \printpercent{\PercentPracticeOpportunitiesTransferableknowledgesecondaryschools} & \DrawPercentageBar{\PercentPracticeOpportunitiesTransferableknowledgesecondaryschools} &  \printpercent{\PercentTeacherOpportunitiesTransferableknowledgesecondaryschools} & \DrawPercentageBar{\PercentTeacherOpportunitiesTransferableknowledgesecondaryschools}\\
 universal &  &  \printpercent{\PercentPracticeOpportunitiesTransferableknowledgegeneral} & \DrawPercentageBar{\PercentPracticeOpportunitiesTransferableknowledgegeneral} &  \printpercent{\PercentTeacherOpportunitiesTransferableknowledgegeneral} & \DrawPercentageBar{\PercentTeacherOpportunitiesTransferableknowledgegeneral}\\

\rowcolor{Gray}
Cross-curricularity &  &  \printpercent{\PercentPracticeOpportunitiesCrosscurricularlearning} & \DrawPercentageBar{\PercentPracticeOpportunitiesCrosscurricularlearning} & \printpercent{\PercentTeacherOpportunitiesCrosscurricularlearning}  & \DrawPercentageBar{\PercentTeacherOpportunitiesCrosscurricularlearning} \\

  \rowcolor{Gray}
Society &  &  \printpercent{\PercentPracticeOpportunitiesSocialimportance} & \DrawPercentageBar{\PercentPracticeOpportunitiesSocialimportance} & \printpercent{\PercentTeacherOpportunitiesSocialimportance} & \DrawPercentageBar{\PercentTeacherOpportunitiesSocialimportance} \\
digitalisation  &  & \printpercent{\PercentPracticeOpportunitiesSocialimportanceincreasingdigitization} & \DrawPercentageBar{\PercentPracticeOpportunitiesSocialimportanceincreasingdigitization} & \printpercent{\PercentTeacherOpportunitiesSocialimportanceincreasingdigitization}  &\DrawPercentageBar{\PercentTeacherOpportunitiesSocialimportanceincreasingdigitization} \\
 student's lifeworld  &  & \printpercent{\PercentPracticeOpportunitiesSocialimportancestudentsslifeworld} & \DrawPercentageBar{\PercentPracticeOpportunitiesSocialimportancestudentsslifeworld} & \printpercent{\PercentTeacherOpportunitiesSocialimportancestudentsslifeworld}  &\DrawPercentageBar{\PercentTeacherOpportunitiesSocialimportancestudentsslifeworld} \\
 
 \rowcolor{Gray}
Early support &  & \printpercent{\PercentPracticeOpportunitiesEarlysupport} & \DrawPercentageBar{\PercentPracticeOpportunitiesEarlysupportgeneral} &\printpercent{\PercentTeacherOpportunitiesEarlysupportgeneral} & \DrawPercentageBar{\PercentTeacherOpportunitiesEarlysupportgeneral} \\
 
 \rowcolor{Gray}
Diversity &  &  \printpercent{\PercentPracticeOpportunitiesDiversity} & \DrawPercentageBar{\PercentPracticeOpportunitiesDiversity} & \printpercent{\PercentTeacherOpportunitiesDiversity}  & \DrawPercentageBar{\PercentTeacherOpportunitiesDiversity} \\
girls &  &\printpercent{\PercentPracticeOpportunitiesDiversitygirls} & \DrawPercentageBar{\PercentPracticeOpportunitiesDiversitygirls}  & \printpercent{\PercentTeacherOpportunitiesDiversitygirls}  & \DrawPercentageBar{\PercentTeacherOpportunitiesDiversitygirls} \\
 alignment &  & \printpercent{\PercentPracticeOpportunitiesDiversityalignmentofstudents}  &  \DrawPercentageBar{\PercentPracticeOpportunitiesDiversityalignmentofstudents} & \printpercent{\PercentTeacherOpportunitiesDiversityalignmentofstudents}  & \DrawPercentageBar{\PercentTeacherOpportunitiesDiversityalignmentofstudents} \\
  girls and boys &  & \printpercent{\PercentPracticeOpportunitiesDiversitybothgirlsandboys}  &  \DrawPercentageBar{\PercentPracticeOpportunitiesDiversitybothgirlsandboys} & \printpercent{\PercentTeacherOpportunitiesDiversitybothgirlsandboys}  & \DrawPercentageBar{\PercentTeacherOpportunitiesDiversitybothgirlsandboys} \\

 \rowcolor{Gray}
 Methods &  & \printpercent{\PercentPracticeOpportunitiesMethods} & \DrawPercentageBar{\PercentPracticeOpportunitiesMethods} & \printpercent{\PercentTeacherOpportunitiesMethods}  & \DrawPercentageBar{\PercentTeacherOpportunitiesMethods}  \\
 
 exploring-discovery & & \printpercent{\PercentPracticeOpportunitiesMethodsexploringdiscoverylearning} & \DrawPercentageBar{\PercentPracticeOpportunitiesMethodsexploringdiscoverylearning}  & \printpercent{\PercentTeacherOpportunitiesMethodsexploringdiscoverylearning} & \DrawPercentageBar{\PercentTeacherOpportunitiesMethodsexploringdiscoverylearning} \\
 
 playful-child-oriented & & \printpercent{\PercentPracticeOpportunitiesMethodsplayfulchildorientedrealization} & \DrawPercentageBar{\PercentPracticeOpportunitiesMethodsplayfulchildorientedrealization}  & \printpercent{\PercentTeacherOpportunitiesMethodsplayfulchildorientedrealization}  & \DrawPercentageBar{ \PercentTeacherOpportunitiesMethodsplayfulchildorientedrealization}  \\
 
 variety & & \printpercent{\PercentPracticeOpportunitiesMethodsvariety} & \DrawPercentageBar{\PercentPracticeOpportunitiesMethodsvariety}  & \printpercent{\PercentTeacherOpportunitiesMethodsvariety}  & \DrawPercentageBar{ \PercentTeacherOpportunitiesMethodsvariety}  \\

\bottomrule
\end{tabular}
}
\end{table}

To answer RQ3 we consider the open question on opportunities
experienced by the \tp (survey 2). \Cref{tab:opportunitiesCodingScheme} shows
the seven identified categories of opportunities with their subcategories and
themes.

\paragraph{Skills acquisition.}
The most frequently mentioned opportunity is the acquisition of skills
(\printpercent{\PercentPracticeOpportunitiesSkillsacquisition}), which is about
building new knowledge and abilities, or about repeating and deepening what has
been learned. It is categorised into metacognitive, cognitive, affective, and
holistic factors. According to the \tpEnd, digital and computational literacy
in particular (\printpercent{\PercentPracticeOpportunitiesSkillsacquisitionholisticdigitalliteracy}, \printpercent{\PercentPracticeOpportunitiesSkillsacquisitionholisticcomputationalliteracy}) as well as a creative component (\printpercent{\PercentPracticeOpportunitiesSkillsacquisitionholisticcreativity}) are fostered through programming.
On a cognitive level, programming is especially important for logical thinking
and problem solving (\printpercent{\PercentPracticeOpportunitiesSkillsacquisitioncognitiveproblemsolvingandlogicalthinking}), whereas it is striking that a small proportion of \tp also
see linguistic support (\printpercent{\PercentPracticeOpportunitiesSkillsacquisitioncognitivelinguisticstimulation}). Starting programming at an early age also offers
opportunities for affective factors, first and foremost the expression of
interest and motivation (\printpercent{\PercentPracticeOpportunitiesSkillsacquisitionaffectiveinterestandmotivation}). In addition to fun (\printpercent{\PercentPracticeOpportunitiesSkillsacquisitionaffectivefun}), active programming in primary
schools can reduce prejudices and fears about programming and computer
science in general (\printpercent{\PercentPracticeOpportunitiesSkillsacquisitionaffectivereductionofprejudices}). Furthermore, self-confidence and independence of the
students are promoted (\printpercent{\PercentPracticeOpportunitiesSkillsacquisitionaffectiveselfconfidence}).

\paragraph{Foundation.}
According to more than a third of the \tpEnd, programming is an important building block for transferring knowledge for the future (\printpercent{\PercentPracticeOpportunitiesTransferableknowledge}). This includes not only skills for the general future life of the students, but also their time at secondary school and later working life: 
\tpquote{65}{Programming is becoming a market in terms of employment and I think teaching children it from a young age will help them develop skills which they could take forward into the future careers and can be built upon at secondary schools.}

\paragraph{Cross-curricularity.}
The opportunity of interdisciplinary learning and teaching while programming is
the third most frequently mentioned category
(\printpercent{\PercentPracticeOpportunitiesCrosscurricularlearning}), which
mainly benefits \tpquote{123}{maths and informatics} or
\tpquote{162}{programming can be integrated across a range of subject areas,
including maths, English, science, geography, history, music and art.}

\paragraph{Society.}
Similar to interdisciplinarity, programming lessons are seen as an essential opportunity to keep pace with the increasing digitalisation of society (\printpercent{\PercentPracticeOpportunitiesSocialimportance}): 
\tpquote{170}{Programming is the foundational language of technology, which is quickly becoming the foundation of society. I think studnets [sic] understanding basic progroaming [sic] is like undertstanding [sic] the basics of a car.}

\paragraph{Early support.}
It is particularly early support, beginning in primary school, that \printpercent{\PercentPracticeOpportunitiesEarlysupport} of the surveyed teachers see as fundamental: 
\tpquote{150}{I think elementary school is the perfect time to give students exposure to the foundations and building blocks of coding. If the seeds can be planted at an early age the [sic] can develop their skills as they progress through school.} 

\paragraph{Diversity.}
Programming might strengthen equal opportunities for all students regardless of  gender or background by reducing mutual reservations and possible stereotypes (\printpercent{\PercentPracticeOpportunitiesDiversity}),
 \tpquote{44}{especially for girls and those children who struggle or are turned off by core subjects} and
 \tpquote{136}{very often, children who are not so academic---and don't do well in other lessons---can really fly in computing lessons.}

\paragraph{Methods.}
There are opportunities for different teaching and learning methods (\printpercent{\PercentPracticeOpportunitiesMethods}), such as 
\tpquote{0}{group work and interaction} or
\tpquote{3}{more active learning}. 

\summary{RQ 3}{The main opportunities of \tp are the development of skills on different levels, interdisciplinarity and a foundation for the future. The early and diversity-related promotion as well as the methodical implementation are also relevant.}

\subsection{RQ 4: Opportunities Perceived by \tt and Strategies for these from \tp}
\label{sec:rq4}

\begin{table}[t]\centering
	\caption{Top ten opportunities of \tt ranked by \tp.\\
	\textmd{\small The \% TP corresponds to the proportion of TP ranking the respective challenge among the three most relevant challenges.}
}
	\vspace{-1em}
	\label{tab:oppsRanking}
	\setlength{\tabcolsep}{0.2em}
	\resizebox{\columnwidth}{!}{
\begin{tabular}{rlp{5.8cm}rlrl}
	\toprule
	Nr. & Category & Opportunity & \% TP &\\
	\midrule

	1 & Skills & Students gain cognitive skills (e.g. problem solving skills).  & \printpercent{\OpportunityPercentThree} & \DrawPercentageBar{\OpportunityPercentThree} \\
	2 & Foundation & Students might acquire knowledge on which they can build on in their future.  & \printpercent{\OpportunityPercentOne} & \DrawPercentageBar{\OpportunityPercentOne}	\\
	3 & 	Skills & Students acquire competencies in terms of digital literacy (e.g. being able to use a computer). & \printpercent{\OpportunityPercentTwo} & \DrawPercentageBar{\OpportunityPercentTwo}\\
	4 & 	Society & Programming and related skills are important in our society (e.g. because of the increasing digitization). & \printpercent{\OpportunityPercentEight} & \DrawPercentageBar{\OpportunityPercentEight}   \\
	5 & Skills & Students acquire positive attitudes towards programming (e.g. increased interest in programming).  & \printpercent{\OpportunityPercentFour} & \DrawPercentageBar{\OpportunityPercentFour}  \\
	6 & Skills & Students are encouraged in their creativity (e.g. by creating an own program).  & \printpercent{\OpportunityPercentNine} & \DrawPercentageBar{\OpportunityPercentNine}  \\
	7 & Skills & Students acquire competencies in terms of computational literacy (e.g. understanding how algorithms work).  & \printpercent{\OpportunityPercentTen} & \DrawPercentageBar{\OpportunityPercentTen}  \\
	8 & Early support & Students are promoted at an early stage.  & \printpercent{\OpportunityPercentSeven} & \DrawPercentageBar{\OpportunityPercentSeven} \\
	9 & Diversity & Boys and especially girls acquire competencies (e.g. wrt. affective aspects such as motivation).  & \printpercent{\OpportunityPercentFive} & \DrawPercentageBar{\OpportunityPercentFive} \\
	10 & Methods & Teaching programming implies new or varied teaching methods.   & \printpercent{\OpportunityPercentSix} & \DrawPercentageBar{\OpportunityPercentSix}   \\

\bottomrule
\end{tabular}
}
\end{table}


To answer RQ4 we consider the perceived opportunities stated by both the \tt (survey 1) and \tp (survey 2), as well as the remaining questions on opportunities and strategies of survey 2. 
\Cref{tab:oppsRanking} shows the ten most common opportunities mentioned by \tt ranked by \tp and \cref{tab:oppsheatmap} includes the proposed strategies by \tp. \Cref{fig:genderLikert} shows the opinions of \tp on gender-homogeneous classes.

\subsubsection{Opportunities Perceived by \tt}
All categories and subcategories are mentioned by both \tt and \tp (\Cref{tab:opportunitiesCodingScheme}). However, the distribution differs which can be explained by significant differences in the subcategories. In the following we discuss these differences.

\paragraph{Skills acquisition.}
In agreement with the \tp (\Cref{tab:opportunitiesCodingScheme}), building new knowledge is the most relevant opportunity (\printpercent{\PercentTeacherOpportunitiesSkillsacquisition}) for the \ttEnd, although the \tt mention it about a quarter more often. 
Both groups see cognitive skills as an essential opportunity.
At the holistic level, the \tt focus on digital and computing literacy as core skills acquired through programming, which differs significantly from the \tp (\pvalue{\PValueOpportunitiesSkillsacquisitionholisticcomputationalliteracy} and \pvalue{\PValueOpportunitiesSkillsacquisitionholisticdigitalliteracy}, respectively). This  points to a misconception of programming~\cite{papadakis2016developing} as many \tt agree that programming will serve  
\ttquote{17}{to encourage children at an early age in the proper use of technology, media, and the Internet.} 

In creativity, both groups see an equal opportunity.
In affective factors, \tt identify similar opportunities as \tpEnd, but they emphasise increasing student interest and motivation and talent (\pvalue{\PValueOpportunitiesSkillsacquisitionaffectiveinterestandmotivation}): 
\ttquote{26}{Particularly at primary school age, one discovers a number of interests, gifts or talents. Accordingly, programming could also be one of them, which means that I can well imagine that a new field of interest will open up for some students'}.
\tp might not share this view due to their practical experience that  students are already very interested in programming and computer science in general, as the survey on the girls' courses (Section \ref{sec:girls}) also indicates.

\paragraph{Foundation.}
Developing skills for the future life is one of the central opportunities for both \tt and \tpEnd. The \tt, in contrast to the \tpEnd, place their focus here on competencies that may be important for the general future (\pvalue{\PValueOpportunitiesTransferableknowledgegeneralfuture}) as well as for secondary school (\pvalue{\PValueOpportunitiesTransferableknowledgesecondaryschools}). Since the teachers already using programming in class have experienced the children's use  and opinions, they may already have more concrete ideas about possible career goals.

\paragraph{Early support.}
Early support is a much more relevant opportunity for beginning teachers than for those already practicing (\pvalue{\PValueOpportunitiesEarlysupportgeneral}), which might be due to the fact that \tp already teach it.

\paragraph{Society.}
The \tt mentioned social relevance about twice as often as the \tp (\pvalue{\PValueOpportunitiesSocialimportanceincreasingdigitization}). In addition to increasing digitalisation, they also referred to the impact on children's lifeworlds, which the \tp did not address at all. \tp may not only see the specific lifeworld reference, but also embed it in a larger societal context.

\paragraph{Methods.}
For the \ttEnd, the methodological didactic application is a much more important opportunity than for the \tp, especially to provide a \ttquote{29}{a change from the `normal' forms of teaching for the children}, thus differing in methods (\pvalue{\PValueOpportunitiesMethodsvariety}) and child-oriented implementation with reducing cognitive load (\pvalue{\PValueOpportunitiesMethodsplayfulchildorientedrealization}). 

\paragraph{Diversity.}
The \tt consider the promotion of diversity as a slightly more relevant opportunity than the \tpEnd. They see opportunities especially for girls (\pvalue{\PValueOpportunitiesDiversitygirls}), whereas the equalisation of all students to one level is seen as similar by both groups.

\paragraph{Cross-curricularity.}
Interdisciplinarity was mentioned by \tt, but significantly less than by \tp (\pvalue{\PValueOpportunitiesCrosscurricularlearninggeneral}), which may be related to their lack of connection factors due to little field experience.

\subsubsection{Strategies for Opportunities}

\begin{table}[t]
\centering
\caption{Strategies of TP for the opportunities of TT.\\
	\textmd{\small The absolute values refer to all mentions and the percentage to the proportion of \tp mentioning the subcategory at least once.}
	} 
	\vspace{-1em}
	\setlength{\tabcolsep}{0.2em}
	\label{tab:oppsheatmap}
	\resizebox{\columnwidth}{!}{
\begin{tabular}{llCCCCCCCCCCCr}
\toprule
Category                                       & Subcategory                  & \multicolumn{1}{c}{1} & \multicolumn{1}{c}{2} & \multicolumn{1}{c}{3} & \multicolumn{1}{c}{4} & \multicolumn{1}{c}{5} & \multicolumn{1}{c}{6} & \multicolumn{1}{c}{7} & \multicolumn{1}{c}{8} & \multicolumn{1}{c}{9} & \multicolumn{1}{c}{10} & \multicolumn{1}{c}{$\sum$ }  & \% TP \\
\midrule
Society and  	& future opportunities     	& 3    & 26   & 10& 19  & 2  & 0  & 1  & 0  & 1  & 0  & 62  &  \printpercent{\StrategyPercentStrategiesforOpportunitiesSocialandprivateimportancefuturejobopportunities} \\
personal relevance                                               	& student's lifeworld          	& 17   & 8   & 5 & 13   & 1  & 1  & 2  & 1  & 1  & 1  & 50  & \printpercent{\StrategyPercentStrategiesforOpportunitiesSocialandprivateimportancestudentslifeworld}  \\
                                               	& relevance of prog. & 3    & 9   & 8 & 18   & 3  & 0  & 3  & 0  & 0  & 0  & 44 &  \printpercent{\StrategyPercentStrategiesforOpportunitiesSocialandprivateimportancerelevanceofprogramming}   \\
						& practical application        & 5    & 10   & 4 & 6    & 1  & 2  & 4  & 0  & 2  & 0  & 34 &  \printpercent{\StrategyPercentStrategiesforOpportunitiesSocialandprivateimportancepracticalapplication}  \\
                                               
Methods                                  	& universal          		& 20   & 13   	& 16& 0   & 5  & 8  & 8  & 3  & 0  & 4  & 77  &  \printpercent{\StrategyPercentStrategiesforOpportunitiesMethodsgeneral}  \\
						& task variety               	  & 13   & 6    	& 7& 4    & 6  & 4  & 2  & 2  & 1  & 1  & 46  &  \printpercent{\StrategyPercentStrategiesforOpportunitiesMethodsactivitieschallengeshomework} \\
                                               	& individual needs            	 & 12   & 2    	& 3 & 1   & 2  & 3  & 1  & 0  & 1  & 0  & 25  &  \printpercent{\StrategyPercentStrategiesforOpportunitiesMethodsindividualneedsfeedbackteamworklearningstrategies}   \\
                                              	 & programs                    	 & 4    & 2    	& 0  & 0  & 1  & 1  & 2  & 1  & 0  & 0  & 11   &   \printpercent{\StrategyPercentStrategiesforOpportunitiesMethodsprograms}\\
                                               
Affective factors                     	& creativity                   		& 5    & 0    	& 0  & 0  & 0  & 21 & 0  & 1  & 1  & 0  & 28   &  \printpercent{\StrategyPercentStrategiesforOpportunitiesAffectivestimulatingstudentscreativity}  \\
						& encouragement    & 3   	 & 6& 0      & 3  & 10 & 3  & 1  & 2  & 2  & 0  & 30   & \printpercent{\StrategyPercentStrategiesforOpportunitiesAffectiveencouragementandenthusiam} \\
                                               	& mindset and inclusion      	& 1    & 7    	& 2 & 3    & 4  & 1  & 0  & 1  & 7  & 2  & 28  & \printpercent{\StrategyPercentStrategiesforOpportunitiesAffectivemindsetandintegration}  \\
                                               	& universal 				& 5    & 1    	& 1& 0    & 12 & 2  & 0  & 1  & 2  & 1  & 25   & \printpercent{\StrategyPercentStrategiesforOpportunitiesAffectiveimprovingaffectivefactorsingeneral}  \\
         					& alternative thinking 	& 5    	& 1& 0      & 0  & 1  & 5  & 1  & 1  & 1  & 0  & 15  &  \printpercent{\StrategyPercentStrategiesforOpportunitiesAffectivestimulatingalternativewaysofthinking}   \\
                                               
Organisational factors              	& requirements                	 & 8    & 12   	& 21& 15  & 4  & 4  & 3  & 1  & 1  & 4  & 73  &  \printpercent{\StrategyPercentStrategiesforOpportunitiesOrganizationalrequirementsliketrainingandequipment} \\
           					& inviting experts   		& 0    & 4    	& 0& 4    & 2  & 0  & 1  & 1  & 0  & 0  & 12   & \printpercent{\StrategyPercentStrategiesforOpportunitiesOrganizationalinvitingexperts} \\
Programming                             &                              		& 8    & 10   	& 12& 9   & 2  & 2  & 5  & 1  & 0  & 0  & 49  & \printpercent{\StrategyPercentStrategiesforOpportunitiesGainingprogrammingskillsgeneral}   \\
Cross-curricularity                      &                             		& 25   & 5    	& 5& 3    & 2  & 2  & 3  & 1  & 0  & 1  & 47   &  \printpercent{\StrategyPercentStrategiesforOpportunitiesCrosscurricularlearninggeneral} \\
Early support                              &                             	 	& 1    & 5    	& 1& 0    & 3  & 0  & 0  & 13 & 0  & 0  & 23  &   \printpercent{\StrategyPercentStrategiesforOpportunitiesEarlysupportgeneral} \\
$\sum$                  &                      		        & 138  & 127  	& 95& 98  & 61 & 59 & 37 & 30 & 20 & 14 & 1358 &  \\

\bottomrule
\end{tabular}
}
\end{table}

The ten most common opportunities mentioned by \tt were ranked regarding their relevance by the \tp (\Cref{tab:oppsRanking}). 
For the three opportunities that the TP ranked most relevant, they were asked to describe a strategy. We explain for each opportunity what strategies are suggested (\Cref{tab:oppsheatmap}).

\paragraph{Opportunity 1: Cognitive Skills} 
The most relevant opportunity, gaining cognitive skills (\printpercent{\OpportunityPercentThree}), can be best promoted with a more interdisciplinary curriculum (\Cref{tab:oppsheatmap}): 
\tpquote{81}{Linking problem solving with other areas of the curriculum eg maths}. 
These potential competencies in turn affect other subjects: 
\tpquote{2}{Learning cognitive skills will help students in other subjects, such as math and english.}

A range of methodological and didactic approaches can also be highly supportive for developing cognitive skills (\Cref{tab:oppsheatmap}).
On the one hand, the variety of tasks is important; for instance, activities like team work, workshops or homework help to better understand programming. In addition, the potential of programming competitions has often been highlighted. 
On the other hand, meeting the individual needs of students is an important component. Thereby the focus is on feedback and the application of suitable learning strategies as one \tp mentioned: 
\tpquote{0}{By differentiating my teaching, e.g., by varying the content and the approaches so as not to exclude [...] the ones who do not find programming extremely related to their needs or learning orientation.} 
In addition, positive reinforcement was often referred to: 
\tpquote{114}{Encourage some kind of reward system for digital superstars.}
Moreover, when the teacher addresses the children's life world, it promotes their cognitive abilities.

\paragraph{Opportunity 2: Foundation} 
Pointing out and actively discussing the many job and career opportunities that exist in computer science has the most beneficial effect on the second most relevant opportunity, the children's future (\printpercent{\OpportunityPercentOne}). This includes explaining and demonstrating all the jobs that use programming (\Cref{tab:oppsheatmap}), and also giving 
\tpquote{121}{examples of people who currently work in programming and what their job involves}.

\paragraph{Opportunity 3: Holistic Skills} 
Organisational factors (\Cref{tab:oppsheatmap}) constitute the basis for students' digital literacy which is the third most relevant opportunity (\printpercent{\OpportunityPercentTwo}). Requirements such as the school's media equipment and a restructuring of the timetable that allows sufficient time for programming lessons play a crucial role. In addition, teacher training can contribute to teaching digital literacy to students.
In order to develop digital literacy at all, a variety of methods should be considered, and particularly the basics and craft of programming must be supported: 
\tpquote{99}{Have them code}. 

\paragraph{Opportunity 4: Society and personal relevance} 
By explaining the relevance of programming and potential job opportunities, the relevance of the subject (\printpercent{\OpportunityPercentEight}) is made more accessible:  
\tpquote{45}{Presentation of various areas of life where computer skills are important and that without them you cannot function in the real world.}

\paragraph{Opportunity 5: Affective Skills} 
Students' positive attitude towards programming (\printpercent{\OpportunityPercentFour}) can be achieved in particular by enthusiastic teachers acting as role models when they
 \tpquote{118}{inspire, empower and engage}. 
 Moreover, teachers should encourage students by, e.g.,
\tpquote{65}{developing motivation, passion, commitment and patience} and
 \tpquote{54}{making it a fun activity}.

\paragraph{Opportunity 6: Creativity}
In order to unleash the potential of creativity  (\printpercent{\OpportunityPercentNine}), \tp suggest that it should be explicitly stimulated by 
\tpquote{12}{encourag[ing] students to go `off piste' and use theor [sic] programming in different ways} or 
\tpquote{143}{open ended tasks which enable children to be more creative with programming e.g. design a game with certain criteria rather than program this traffic light}.

\paragraph{Opportunity 7: Holistic skills} 
Computational literacy can be promoted through different methodological approaches (\Cref{tab:oppsheatmap}), e.g., a gradual and playful implementation was emphasised by the \tp: 
\tpquote{136}{Begin with simple algorithms/instructions they are already able to understand and follow, such as food recipes or instructions for getting up in the morning. Build on those. Card sorting activities can help support those who struggle with ordering instructions.}

\paragraph{Opportunity 8: Early support} 
Rather intuitively, early promotion can be met with precisely such an approach, since 
\tpquote{62}{the earlier children explore the basics of coding, the more easily they will be able to learn, understand, and apply coding later in life.}

\paragraph{Opportunity 9: Diversity} 
Diversity can best be promoted by positively changing the attitudes of parents and teachers (\Cref{tab:oppsheatmap}), e.g., by doing more public relations work such as students giving 
\tpquote{102}{a presentation about programming, to parents or during an assembly} 
and by establishing an inclusive climate, where the school can also be a place where equal opportunities are created, since 
\tpquote{174}{many [students] do not have access to computers at home. So it is a must to provide the opportunity at school.}

\paragraph{Opportunity 10: Methods} 
In order to apply the types of teaching that programming education requires, teachers need to be taught how to use \tpquote{0}{new tools that facilitate both teaching (approaches, load, differentiation of content and process) and learning [...].}

\subsubsection{Gender Homogeneous Groups}
\label{sec:girls}
\begin{figure}[tb]
	\centering
	\includegraphics[width=\columnwidth]{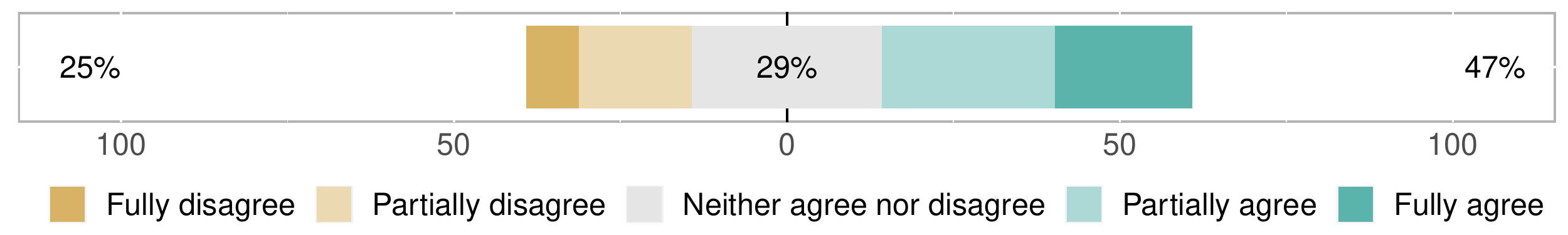}
	\vspace{-2em}
	\caption{\label{fig:genderLikert} Opinions of TP on gender-homogeneous classes.}
\end{figure}

Both \tt (\printpercent{\PercentTeacherOpportunitiesDiversitygirls}) and \tp (\printpercent{\PercentPracticeOpportunitiesDiversitygirls}) see the promotion of girls as an opportunity for programming in primary schools, and in general, diversity in computer science has been gaining importance in recent years~\cite{albusays2021diversity}. One way to get girls interested in programming might be to offer special courses just for girls~\cite{zhan2015effects}. \Cref{fig:genderLikert} shows that the majority of \tp consider such courses to be positive or are neutral toward them. 

Many teachers report that they experienced few to no affective (\printpercent{\StrategyPercentHomogeneousCoursesSimilaritiesaffective}) and cognitive (\printpercent{\StrategyPercentHomogeneousCoursesSimilaritiescognitive}) differences between the genders in primary school, although there may be preferences thematically~\cite{grassl2021data}, as
 \tpquote{73}{boys are more interested in `boy' type games and programs. The girls do enjoy minecraft which is more neutral.}

While teachers see it as important to promote girls, they are also skeptical since mixed courses give the opportunity to impact the working environment in the long-term:
\tpquote{189}{In order to break down the stereotyping which exists around IT and other technologically related career paths, I believe that they should be taught in mixed classes where girls are able to show off how skilled they are at coding. By doing so we will have a new generation of men that respect the programming abilities of their opposite gender.}

Gender bias may also be present at the teacher level and should be included: 
\tpquote{136}{I think teacher awareness and support is much more relevant. There may be biases towards boys doing better, but that will not always be the case and the teacher needs to push all pupils to the best of their abilities, whilst keeping in mind the biases that exist in their class or school.} 
Above all, interaction between the genders should be encouraged:
\tpquote{7}{cooperation instead of antipathy}


Since diversity has several aspects besides gender, some teachers see the bigger picture in pointing out the importance of other underrepresented groups: 
\tpquote{44}{In my district, there is a focus on raising the attainment of children from poorer families and this would be a bigger concern than the gender attainment gap.}

Overall, teachers see the support of girls as very relevant, but are ambivalent about whether homogeneous courses are a suitable implementation for this, in-line with current public discourse~\cite{albusays2021diversity}.

\summary{RQ 4}{The main opportunities of \tt are similar to those of the \tpEnd, but \tt emphasise early support, society and methods. The most conducive strategies of \tp are various methods, interdisciplinarity, and demonstration of the relevance of programming. 
Many \tp are positive about promoting girls through extra courses for girls, but see a danger of further stereotyping. 
}

\section{Discussion}
\label{sec:discussion}

%
Considering challenges in conjunction with opportunities, some overarching misconceptions of \tt and strategies of \tp are recognisable.
%
For example, \tt seem to overestimate the importance of students' digital literacy. This might be related to even in-service teachers having difficulties regarding terms such as `programming' or `computational thinking' \cite{corradini2018investigation,munasinghe2021teachers}.
The surveyed \tp, in contrast, focused on more general cognitive skills such as reasoning and problem solving, which are strongly associated with CT~\cite{selby2014refining}. This goes along with the idea that programming should not be taught as an independent skill but rather be used as an approach to promote CT~\cite{munasinghe2021teachers}. Teacher training should therefore explain programming as a vehicle to promote ways of thinking rather than detached skills.

Teacher training should provide strategies to use programming to teach computational thinking concepts. When comparing the strategies for the associated challenges and opportunities, \didacticconsiderations is emphasised (\cref{tab:challheatmap,tab:oppsheatmap}). \tp suggest, e.g., to consider individual needs but also interdisciplinary teaching, as cross-curricular teaching supports dealing with different challenges and opportunities and moreover, is even an opportunity itself. Likewise, the Computer Science Teachers Association and International Society for Technology in Education (ISTE) points out that CT also needs to be taught across multiple curricular disciplines~\cite{ray2020perceptions} as one teacher suggests:
\tpquote{119}{Put programming as a stand alone lesson to teach skills but also inlude [sic] it in other lessons to show how versatile it is and to link it to real life problems.}
In particular, programming can refer to selected contents of other subjects and thus establish a synergy of different learning areas. Conceivable learning effects for a simple game in, e.g., \Scratch would be as follows: Math (coordinate system, angles, mirroring), Languages (syntax, verbal and nonverbal signals, rules), Physics (movement, gravitation, distance and time measurement), Music (gaits, dance, sounds and noises), Art (colours) and general studies (senses, food, clothing, dances, rituals and festivals).

Such a cognitively reduced, playful and interdisciplinary approach is not only useful for children~\cite{bers2019coding}, but also for teachers. Especially for the \tt, \Scratch courses at university can reduce fears of contact through self-experience~\cite{yukselturk2017investigation} and also convey the competencies that are relevant for their future teaching practice~\cite{yeni2020or}.

\section{Conclusions and Future Work}
\label{conclusions}


In order to bridge the gap between concept and practice and provide a realistic picture of programming education for teachers, we surveyed 200 \tp and 97 \tt regarding their opinions on the challenges and opportunities of teaching programming in primary schools.
Overall, our results show that, despite some challenges, the majority of both \tt and \tp support the introduction of programming into primary schools and see it as valuable. 

%
\tp and \tt perceive challenges regarding the school, students, teachers, programming, government and parents, whereas \tt set a different focus. To counter challenges, \tp consider especially teacher training and \didacticconsiderations as useful. Their anticipation of the help provided by automated analysis tools provides reinforcement for the research, development, and evaluation of such tools.
According to \tp, programming classes also provide opportunities, especially to develop skills on different levels which children will need for their future life. Similar to the challenges, the \tt and \tp identify similar opportunities but emphasise them differently.
Programming might promote girls into computer science, though 
different course designs should be reviewed in field studies.
The many opportunities will only yield benefits if programming is also embedded in the curriculum, and if the strategies identified in this study are applied in teacher training and evaluated in practice.

\vspace{-0.3em}
\begin{acks}
\vspace{-0.3em}This work is supported by the Federal Ministry of Education and Research
through project ``primary::programming'' (01JA2021) as
part of the ``Qualitätsoffensive Lehrerbildung'', a joint initiative of the
Federal Government and the Länder. The authors are responsible for the content
of this publication.
\end{acks}

 \bibliographystyle{ACM-Reference-Format}
 \bibliography{references}


\begin{thebibliography}{41}


\ifx \showCODEN    \undefined \def \showCODEN     #1{\unskip}     \fi
\ifx \showDOI      \undefined \def \showDOI       #1{#1}\fi
\ifx \showISBNx    \undefined \def \showISBNx     #1{\unskip}     \fi
\ifx \showISBNxiii \undefined \def \showISBNxiii  #1{\unskip}     \fi
\ifx \showISSN     \undefined \def \showISSN      #1{\unskip}     \fi
\ifx \showLCCN     \undefined \def \showLCCN      #1{\unskip}     \fi
\ifx \shownote     \undefined \def \shownote      #1{#1}          \fi
\ifx \showarticletitle \undefined \def \showarticletitle #1{#1}   \fi
\ifx \showURL      \undefined \def \showURL       {\relax}        \fi
\providecommand\bibfield[2]{#2}
\providecommand\bibinfo[2]{#2}
\providecommand\natexlab[1]{#1}
\providecommand\showeprint[2][]{arXiv:#2}

\bibitem[\protect\citeauthoryear{Albusays, Bjorn, Dabbish, Ford, Murphy-Hill,
  Serebrenik, and Storey}{Albusays et~al\mbox{.}}{2021}]%
        {albusays2021diversity}
\bibfield{author}{\bibinfo{person}{K. Albusays}, \bibinfo{person}{P. Bjorn},
  \bibinfo{person}{L. Dabbish}, \bibinfo{person}{D. Ford}, \bibinfo{person}{E.
  Murphy-Hill}, \bibinfo{person}{A. Serebrenik}, {and} \bibinfo{person}{M.-A.
  Storey}.} \bibinfo{year}{2021}\natexlab{}.
\newblock \showarticletitle{The diversity crisis in software development}.
\newblock \bibinfo{journal}{\emph{IEEE Software}} \bibinfo{volume}{38},
  \bibinfo{number}{2} (\bibinfo{year}{2021}), \bibinfo{pages}{19--25}.
\newblock


\bibitem[\protect\citeauthoryear{Alves, Von~Wangenheim, and Hauck}{Alves
  et~al\mbox{.}}{2019}]%
        {alves2019approaches}
\bibfield{author}{\bibinfo{person}{N.~D.~C. Alves}, \bibinfo{person}{C.~G.
  Von~Wangenheim}, {and} \bibinfo{person}{J.~C. Hauck}.}
  \bibinfo{year}{2019}\natexlab{}.
\newblock \showarticletitle{Approaches to assess computational thinking
  competences based on code analysis in K-12 education: A systematic mapping
  study}.
\newblock \bibinfo{journal}{\emph{Informatics in Education}}
  \bibinfo{volume}{18}, \bibinfo{number}{1} (\bibinfo{year}{2019}),
  \bibinfo{pages}{17}.
\newblock


\bibitem[\protect\citeauthoryear{Arf{\'e}, Vardanega, and Ronconi}{Arf{\'e}
  et~al\mbox{.}}{2020}]%
        {arfe2020effects}
\bibfield{author}{\bibinfo{person}{B. Arf{\'e}}, \bibinfo{person}{T.
  Vardanega}, {and} \bibinfo{person}{L. Ronconi}.}
  \bibinfo{year}{2020}\natexlab{}.
\newblock \showarticletitle{The effects of coding on children's planning and
  inhibition skills}.
\newblock \bibinfo{journal}{\emph{Computers \& Education}}
  \bibinfo{volume}{148} (\bibinfo{year}{2020}), \bibinfo{pages}{103807}.
\newblock


\bibitem[\protect\citeauthoryear{Ausiku and Matthee}{Ausiku and
  Matthee}{2020}]%
        {ausiku2020preparing}
\bibfield{author}{\bibinfo{person}{M. Ausiku} {and} \bibinfo{person}{M.
  Matthee}.} \bibinfo{year}{2020}\natexlab{}.
\newblock \showarticletitle{Preparing Primary School Teachers for Teaching
  Computational Thinking: A Systematic Review}.
\newblock \bibinfo{journal}{\emph{Learning Technologies and Systems}}
  (\bibinfo{year}{2020}), \bibinfo{pages}{202--213}.
\newblock


\bibitem[\protect\citeauthoryear{Bell, Witten, and Fellows}{Bell
  et~al\mbox{.}}{2015}]%
        {bell2015cs}
\bibfield{author}{\bibinfo{person}{T. Bell}, \bibinfo{person}{I. Witten}, {and}
  \bibinfo{person}{M. Fellows}.} \bibinfo{year}{2015}\natexlab{}.
\newblock \showarticletitle{CS Unplugged: An enrichment and extension programme
  for primary-aged students}.
\newblock  (\bibinfo{year}{2015}).
\newblock


\bibitem[\protect\citeauthoryear{Bergman}{Bergman}{2010}]%
        {bergman2010hermeneutic}
\bibfield{author}{\bibinfo{person}{M.~M. Bergman}.}
  \bibinfo{year}{2010}\natexlab{}.
\newblock \showarticletitle{Hermeneutic content analysis: Textual and
  audiovisual analyses within a mixed methods framework}.
\newblock \bibinfo{journal}{\emph{SAGE Handbook of Mixed Methods in Social and
  Behavioral Research. Thousand Oaks, SAGE}} (\bibinfo{year}{2010}),
  \bibinfo{pages}{379--396}.
\newblock


\bibitem[\protect\citeauthoryear{Bers, Gonz{\'a}lez-Gonz{\'a}lez, and
  Armas-Torres}{Bers et~al\mbox{.}}{2019}]%
        {bers2019coding}
\bibfield{author}{\bibinfo{person}{M.~U. Bers}, \bibinfo{person}{C.
  Gonz{\'a}lez-Gonz{\'a}lez}, {and} \bibinfo{person}{M.~B. Armas-Torres}.}
  \bibinfo{year}{2019}\natexlab{}.
\newblock \showarticletitle{Coding as a playground: Promoting positive learning
  experiences in childhood classrooms}.
\newblock \bibinfo{journal}{\emph{Computers \& Education}}
  \bibinfo{volume}{138} (\bibinfo{year}{2019}), \bibinfo{pages}{130--145}.
\newblock


\bibitem[\protect\citeauthoryear{{\c{C}}iftci and Bildiren}{{\c{C}}iftci and
  Bildiren}{2020}]%
        {cciftci2020effect}
\bibfield{author}{\bibinfo{person}{S. {\c{C}}iftci} {and} \bibinfo{person}{A.
  Bildiren}.} \bibinfo{year}{2020}\natexlab{}.
\newblock \showarticletitle{The effect of coding courses on the cognitive
  abilities and problem-solving skills of preschool children}.
\newblock \bibinfo{journal}{\emph{Computer science education}}
  \bibinfo{volume}{30}, \bibinfo{number}{1} (\bibinfo{year}{2020}),
  \bibinfo{pages}{3--21}.
\newblock


\bibitem[\protect\citeauthoryear{Corradini, Lodi, and Nardelli}{Corradini
  et~al\mbox{.}}{2018}]%
        {corradini2018investigation}
\bibfield{author}{\bibinfo{person}{I. Corradini}, \bibinfo{person}{M. Lodi},
  {and} \bibinfo{person}{E. Nardelli}.} \bibinfo{year}{2018}\natexlab{}.
\newblock \showarticletitle{An investigation of Italian primary school
  teachers’ view on coding and programming}. In
  \bibinfo{booktitle}{\emph{International Conference on Informatics in Schools:
  Situation, Evolution, and Perspectives}}. Springer,
  \bibinfo{pages}{228--243}.
\newblock


\bibitem[\protect\citeauthoryear{Dagiene, Mannila, Poranen, Rolandsson, and
  S{\"o}derhjelm}{Dagiene et~al\mbox{.}}{2014}]%
        {dagiene2014students}
\bibfield{author}{\bibinfo{person}{V. Dagiene}, \bibinfo{person}{L. Mannila},
  \bibinfo{person}{T. Poranen}, \bibinfo{person}{L. Rolandsson}, {and}
  \bibinfo{person}{P. S{\"o}derhjelm}.} \bibinfo{year}{2014}\natexlab{}.
\newblock \showarticletitle{Students' performance on programming-related tasks
  in an informatics contest in Finland, Sweden and Lithuania}. In
  \bibinfo{booktitle}{\emph{Proceedings of the 2014 conference on Innovation \&
  technology in computer science education}}. \bibinfo{pages}{153--158}.
\newblock


\bibitem[\protect\citeauthoryear{Ertmer and Ottenbreit-Leftwich}{Ertmer and
  Ottenbreit-Leftwich}{2010}]%
        {ertmer2010teacher}
\bibfield{author}{\bibinfo{person}{P.~A. Ertmer} {and} \bibinfo{person}{A.~T.
  Ottenbreit-Leftwich}.} \bibinfo{year}{2010}\natexlab{}.
\newblock \showarticletitle{Teacher technology change: How knowledge,
  confidence, beliefs, and culture intersect}.
\newblock \bibinfo{journal}{\emph{Journal of research on Technology in
  Education}} \bibinfo{volume}{42}, \bibinfo{number}{3} (\bibinfo{year}{2010}),
  \bibinfo{pages}{255--284}.
\newblock


\bibitem[\protect\citeauthoryear{Fesakis and Serafeim}{Fesakis and
  Serafeim}{2009}]%
        {fesakis2009influence}
\bibfield{author}{\bibinfo{person}{G. Fesakis} {and} \bibinfo{person}{K.
  Serafeim}.} \bibinfo{year}{2009}\natexlab{}.
\newblock \showarticletitle{Influence of the familiarization with" scratch" on
  future teachers' opinions and attitudes about programming and ICT in
  education}.
\newblock \bibinfo{journal}{\emph{Acm SIGCSE Bulletin}} \bibinfo{volume}{41},
  \bibinfo{number}{3} (\bibinfo{year}{2009}), \bibinfo{pages}{258--262}.
\newblock


\bibitem[\protect\citeauthoryear{Girvan, Conneely, and Tangney}{Girvan
  et~al\mbox{.}}{2016}]%
        {girvan2016extending}
\bibfield{author}{\bibinfo{person}{C. Girvan}, \bibinfo{person}{C. Conneely},
  {and} \bibinfo{person}{B. Tangney}.} \bibinfo{year}{2016}\natexlab{}.
\newblock \showarticletitle{Extending experiential learning in teacher
  professional development}.
\newblock \bibinfo{journal}{\emph{TEACH TEACH EDUC}}  \bibinfo{volume}{58}
  (\bibinfo{year}{2016}), \bibinfo{pages}{129--139}.
\newblock


\bibitem[\protect\citeauthoryear{Gra{\ss}l, Geldreich, and Fraser}{Gra{\ss}l
  et~al\mbox{.}}{2021}]%
        {grassl2021data}
\bibfield{author}{\bibinfo{person}{I. Gra{\ss}l}, \bibinfo{person}{K.
  Geldreich}, {and} \bibinfo{person}{G. Fraser}.}
  \bibinfo{year}{2021}\natexlab{}.
\newblock \showarticletitle{Data-driven Analysis of Gender Differences and
  Similarities in Scratch Programs}. In \bibinfo{booktitle}{\emph{Proceedings
  of the 16th Workshop in Primary and Secondary Computing Education (WiPSCE)}}.
  \bibinfo{publisher}{ACM}.
\newblock
\newblock
\shownote{To appear.}


\bibitem[\protect\citeauthoryear{Greifenstein, Oberm{\"u}ller, Wasmeier, Heuer,
  and Fraser}{Greifenstein et~al\mbox{.}}{2021}]%
        {greifenstein2021effects}
\bibfield{author}{\bibinfo{person}{L. Greifenstein}, \bibinfo{person}{F.
  Oberm{\"u}ller}, \bibinfo{person}{E. Wasmeier}, \bibinfo{person}{U. Heuer},
  {and} \bibinfo{person}{G. Fraser}.} \bibinfo{year}{2021}\natexlab{}.
\newblock \showarticletitle{Effects of Hints on Debugging Scratch Programs: An
  Empirical Study with Primary School Teachers in Training}. In
  \bibinfo{booktitle}{\emph{Proceedings of the 16th Workshop in Primary and
  Secondary Computing Education (WiPSCE)}}. \bibinfo{publisher}{ACM}.
\newblock
\newblock
\shownote{To appear.}


\bibitem[\protect\citeauthoryear{Hattie}{Hattie}{2008}]%
        {hattie2008visible}
\bibfield{author}{\bibinfo{person}{J. Hattie}.}
  \bibinfo{year}{2008}\natexlab{}.
\newblock \bibinfo{booktitle}{\emph{Visible learning: A synthesis of over 800
  meta-analyses relating to achievement}}.
\newblock \bibinfo{publisher}{routledge}.
\newblock


\bibitem[\protect\citeauthoryear{Heintz, Mannila, and Färnqvist}{Heintz
  et~al\mbox{.}}{2016}]%
        {heintz2016}
\bibfield{author}{\bibinfo{person}{F. Heintz}, \bibinfo{person}{L. Mannila},
  {and} \bibinfo{person}{T. Färnqvist}.} \bibinfo{year}{2016}\natexlab{}.
\newblock \showarticletitle{A review of models for introducing computational
  thinking, computer science and computing in K-12 education}. In
  \bibinfo{booktitle}{\emph{FIE '16}}. \bibinfo{pages}{1--9}.
\newblock


\bibitem[\protect\citeauthoryear{Israel, Pearson, Tapia, Wherfel, and
  Reese}{Israel et~al\mbox{.}}{2015}]%
        {israel2015supporting}
\bibfield{author}{\bibinfo{person}{M. Israel}, \bibinfo{person}{J.~N. Pearson},
  \bibinfo{person}{T. Tapia}, \bibinfo{person}{Q.~M. Wherfel}, {and}
  \bibinfo{person}{G. Reese}.} \bibinfo{year}{2015}\natexlab{}.
\newblock \showarticletitle{Supporting all learners in school-wide
  computational thinking: A cross-case qualitative analysis}.
\newblock \bibinfo{journal}{\emph{Computers \& Education}}
  \bibinfo{volume}{82} (\bibinfo{year}{2015}), \bibinfo{pages}{263--279}.
\newblock


\bibitem[\protect\citeauthoryear{Kafai, Proctor, and Lui}{Kafai
  et~al\mbox{.}}{2020}]%
        {kafai2020theory}
\bibfield{author}{\bibinfo{person}{Y. Kafai}, \bibinfo{person}{C. Proctor},
  {and} \bibinfo{person}{D. Lui}.} \bibinfo{year}{2020}\natexlab{}.
\newblock \showarticletitle{From theory bias to theory dialogue: embracing
  cognitive, situated, and critical framings of computational thinking in K-12
  CS education}.
\newblock \bibinfo{journal}{\emph{ACM Inroads}} \bibinfo{volume}{11},
  \bibinfo{number}{1} (\bibinfo{year}{2020}), \bibinfo{pages}{44--53}.
\newblock


\bibitem[\protect\citeauthoryear{Koulouri, Lauria, and Macredie}{Koulouri
  et~al\mbox{.}}{2014}]%
        {koulouri2014teaching}
\bibfield{author}{\bibinfo{person}{T. Koulouri}, \bibinfo{person}{S. Lauria},
  {and} \bibinfo{person}{R.~D. Macredie}.} \bibinfo{year}{2014}\natexlab{}.
\newblock \showarticletitle{Teaching introductory programming: A quantitative
  evaluation of different approaches}.
\newblock \bibinfo{journal}{\emph{ACM Transactions on Computing Education
  (TOCE)}} \bibinfo{volume}{14}, \bibinfo{number}{4} (\bibinfo{year}{2014}),
  \bibinfo{pages}{1--28}.
\newblock


\bibitem[\protect\citeauthoryear{Larke}{Larke}{2019}]%
        {larke2019agentic}
\bibfield{author}{\bibinfo{person}{L.~R. Larke}.}
  \bibinfo{year}{2019}\natexlab{}.
\newblock \showarticletitle{Agentic neglect: Teachers as gatekeepers of
  England’s national computing curriculum}.
\newblock \bibinfo{journal}{\emph{BJET}} \bibinfo{volume}{50},
  \bibinfo{number}{3} (\bibinfo{year}{2019}), \bibinfo{pages}{1137--1150}.
\newblock


\bibitem[\protect\citeauthoryear{Mason and Rich}{Mason and Rich}{2019}]%
        {mason2019preparing}
\bibfield{author}{\bibinfo{person}{S.~L. Mason} {and} \bibinfo{person}{P.~J.
  Rich}.} \bibinfo{year}{2019}\natexlab{}.
\newblock \showarticletitle{Preparing elementary school teachers to teach
  computing, coding, and computational thinking}.
\newblock \bibinfo{journal}{\emph{Contemporary Issues in Technology and Teacher
  Education}} \bibinfo{volume}{19}, \bibinfo{number}{4} (\bibinfo{year}{2019}),
  \bibinfo{pages}{790--824}.
\newblock


\bibitem[\protect\citeauthoryear{Michaeli}{Michaeli}{2021}]%
        {michaeli2021debugging}
\bibfield{author}{\bibinfo{person}{T. Michaeli}.}
  \bibinfo{year}{2021}\natexlab{}.
\newblock \emph{\bibinfo{title}{Debugging im Informatikunterricht}}.
\newblock \bibinfo{thesistype}{Ph.D. Dissertation}.
\newblock


\bibitem[\protect\citeauthoryear{Munasinghe, Bell, and Robins}{Munasinghe
  et~al\mbox{.}}{2021}]%
        {munasinghe2021teachers}
\bibfield{author}{\bibinfo{person}{B. Munasinghe}, \bibinfo{person}{T. Bell},
  {and} \bibinfo{person}{A. Robins}.} \bibinfo{year}{2021}\natexlab{}.
\newblock \showarticletitle{Teachers’ understanding of technical terms in a
  Computational Thinking curriculum}. In \bibinfo{booktitle}{\emph{ACE}}.
  \bibinfo{pages}{106--114}.
\newblock


\bibitem[\protect\citeauthoryear{Oberm{\"u}ller, Bloch, Greifenstein, Heuer,
  and Fraser}{Oberm{\"u}ller et~al\mbox{.}}{2021a}]%
        {obermuller2021code}
\bibfield{author}{\bibinfo{person}{F. Oberm{\"u}ller}, \bibinfo{person}{L.
  Bloch}, \bibinfo{person}{L. Greifenstein}, \bibinfo{person}{U. Heuer}, {and}
  \bibinfo{person}{G. Fraser}.} \bibinfo{year}{2021}\natexlab{a}.
\newblock \showarticletitle{Code Perfumes: Reporting Good Code to Encourage
  Learners}. In \bibinfo{booktitle}{\emph{Proceedings of the 16th Workshop in
  Primary and Secondary Computing Education (WiPSCE)}}.
  \bibinfo{publisher}{ACM}.
\newblock
\newblock
\shownote{To appear.}


\bibitem[\protect\citeauthoryear{Oberm{\"u}ller, Heuer, and
  Fraser}{Oberm{\"u}ller et~al\mbox{.}}{2021b}]%
        {obermuller2021guiding}
\bibfield{author}{\bibinfo{person}{F. Oberm{\"u}ller}, \bibinfo{person}{U.
  Heuer}, {and} \bibinfo{person}{G. Fraser}.} \bibinfo{year}{2021}\natexlab{b}.
\newblock \showarticletitle{Guiding Next-Step Hint Generation Using Automated
  Tests}. In \bibinfo{booktitle}{\emph{Proceedings of the 26th ACM Conference
  on Innovation and Technology in Computer Science Education V. 1}}.
  \bibinfo{pages}{220--226}.
\newblock


\bibitem[\protect\citeauthoryear{Papadakis, Kalogiannakis, and
  Zaranis}{Papadakis et~al\mbox{.}}{2016}]%
        {papadakis2016developing}
\bibfield{author}{\bibinfo{person}{S. Papadakis}, \bibinfo{person}{M.
  Kalogiannakis}, {and} \bibinfo{person}{N. Zaranis}.}
  \bibinfo{year}{2016}\natexlab{}.
\newblock \showarticletitle{Developing fundamental programming concepts and
  computational thinking with ScratchJr in preschool education: a case study}.
\newblock \bibinfo{journal}{\emph{IJMLO}} \bibinfo{volume}{10},
  \bibinfo{number}{3} (\bibinfo{year}{2016}), \bibinfo{pages}{187--202}.
\newblock


\bibitem[\protect\citeauthoryear{Ray, Rogers, and Hocutt}{Ray
  et~al\mbox{.}}{2020}]%
        {ray2020perceptions}
\bibfield{author}{\bibinfo{person}{B.~B. Ray}, \bibinfo{person}{R.~R. Rogers},
  {and} \bibinfo{person}{M.~M. Hocutt}.} \bibinfo{year}{2020}\natexlab{}.
\newblock \showarticletitle{Perceptions of non-STEM discipline teachers on
  coding as a teaching and learning tool: what are the possibilities?}
\newblock \bibinfo{journal}{\emph{Journal of Digital Learning in Teacher
  Education}} \bibinfo{volume}{36}, \bibinfo{number}{1} (\bibinfo{year}{2020}),
  \bibinfo{pages}{19--31}.
\newblock


\bibitem[\protect\citeauthoryear{Resnick, Maloney, Monroy-Hern{\'a}ndez, Rusk,
  Eastmond, Brennan, Millner, Rosenbaum, Silver, Silverman,
  et~al\mbox{.}}{Resnick et~al\mbox{.}}{2009}]%
        {resnick2009scratch}
\bibfield{author}{\bibinfo{person}{M. Resnick}, \bibinfo{person}{J. Maloney},
  \bibinfo{person}{A. Monroy-Hern{\'a}ndez}, \bibinfo{person}{N. Rusk},
  \bibinfo{person}{E. Eastmond}, \bibinfo{person}{K. Brennan},
  \bibinfo{person}{A. Millner}, \bibinfo{person}{E. Rosenbaum},
  \bibinfo{person}{J. Silver}, \bibinfo{person}{B. Silverman}, {et~al\mbox{.}}}
  \bibinfo{year}{2009}\natexlab{}.
\newblock \showarticletitle{Scratch: programming for all}.
\newblock \bibinfo{journal}{\emph{Commun. ACM}} \bibinfo{volume}{52},
  \bibinfo{number}{11} (\bibinfo{year}{2009}), \bibinfo{pages}{60--67}.
\newblock


\bibitem[\protect\citeauthoryear{Rich, Browning, Perkins, Shoop, Yoshikawa, and
  Belikov}{Rich et~al\mbox{.}}{2019}]%
        {rich2019coding}
\bibfield{author}{\bibinfo{person}{P.~J. Rich}, \bibinfo{person}{S.~F.
  Browning}, \bibinfo{person}{M. Perkins}, \bibinfo{person}{T. Shoop},
  \bibinfo{person}{E. Yoshikawa}, {and} \bibinfo{person}{O.~M. Belikov}.}
  \bibinfo{year}{2019}\natexlab{}.
\newblock \showarticletitle{Coding in K-8: International trends in teaching
  elementary/primary computing}.
\newblock \bibinfo{journal}{\emph{TechTrends}} \bibinfo{volume}{63},
  \bibinfo{number}{3} (\bibinfo{year}{2019}), \bibinfo{pages}{311--329}.
\newblock


\bibitem[\protect\citeauthoryear{Ryder}{Ryder}{2015}]%
        {ryder2015being}
\bibfield{author}{\bibinfo{person}{J. Ryder}.} \bibinfo{year}{2015}\natexlab{}.
\newblock \showarticletitle{Being professional: accountability and authority in
  teachers’ responses to science curriculum reform}.
\newblock \bibinfo{journal}{\emph{Stud Sci Educ}} \bibinfo{volume}{51},
  \bibinfo{number}{1} (\bibinfo{year}{2015}), \bibinfo{pages}{87--120}.
\newblock


\bibitem[\protect\citeauthoryear{S{\'a}ez-L{\'o}pez, del Olmo-Mu{\~n}oz,
  Gonz{\'a}lez-Calero, and C{\'o}zar-Guti{\'e}rrez}{S{\'a}ez-L{\'o}pez
  et~al\mbox{.}}{2020}]%
        {saez2020exploring}
\bibfield{author}{\bibinfo{person}{J.~M. S{\'a}ez-L{\'o}pez},
  \bibinfo{person}{J. del Olmo-Mu{\~n}oz}, \bibinfo{person}{J.~A.
  Gonz{\'a}lez-Calero}, {and} \bibinfo{person}{R. C{\'o}zar-Guti{\'e}rrez}.}
  \bibinfo{year}{2020}\natexlab{}.
\newblock \showarticletitle{Exploring the Effect of Training in Visual Block
  Programming for Preservice Teachers}.
\newblock \bibinfo{journal}{\emph{Multimodal Technologies and Interaction}}
  \bibinfo{volume}{4}, \bibinfo{number}{3} (\bibinfo{year}{2020}),
  \bibinfo{pages}{65}.
\newblock


\bibitem[\protect\citeauthoryear{Selby and Woollard}{Selby and
  Woollard}{2014}]%
        {selby2014refining}
\bibfield{author}{\bibinfo{person}{C. Selby} {and} \bibinfo{person}{J.
  Woollard}.} \bibinfo{year}{2014}\natexlab{}.
\newblock \showarticletitle{Refining an understanding of computational
  thinking}.
\newblock  (\bibinfo{year}{2014}).
\newblock


\bibitem[\protect\citeauthoryear{Sentance and Csizmadia}{Sentance and
  Csizmadia}{2017}]%
        {sentance2017computing}
\bibfield{author}{\bibinfo{person}{S. Sentance} {and} \bibinfo{person}{A.
  Csizmadia}.} \bibinfo{year}{2017}\natexlab{}.
\newblock \showarticletitle{Computing in the curriculum: Challenges and
  strategies from a teacher’s perspective}.
\newblock \bibinfo{journal}{\emph{Educ Inform Tech}} \bibinfo{volume}{22},
  \bibinfo{number}{2} (\bibinfo{year}{2017}), \bibinfo{pages}{469--495}.
\newblock


\bibitem[\protect\citeauthoryear{Tuomi, Multisilta, Saarikoski, and
  Suominen}{Tuomi et~al\mbox{.}}{2018}]%
        {tuomi2018coding}
\bibfield{author}{\bibinfo{person}{P. Tuomi}, \bibinfo{person}{J. Multisilta},
  \bibinfo{person}{P. Saarikoski}, {and} \bibinfo{person}{J. Suominen}.}
  \bibinfo{year}{2018}\natexlab{}.
\newblock \showarticletitle{Coding skills as a success factor for a society}.
\newblock \bibinfo{journal}{\emph{Educ Inform Tech}} \bibinfo{volume}{23},
  \bibinfo{number}{1} (\bibinfo{year}{2018}), \bibinfo{pages}{419--434}.
\newblock


\bibitem[\protect\citeauthoryear{Vee}{Vee}{2017}]%
        {vee2017coding}
\bibfield{author}{\bibinfo{person}{A. Vee}.} \bibinfo{year}{2017}\natexlab{}.
\newblock \bibinfo{booktitle}{\emph{Coding literacy: How computer programming
  is changing writing}}.
\newblock \bibinfo{publisher}{Mit Press}.
\newblock


\bibitem[\protect\citeauthoryear{Vinnervik}{Vinnervik}{2020}]%
        {vinnervik2020implementing}
\bibfield{author}{\bibinfo{person}{P. Vinnervik}.}
  \bibinfo{year}{2020}\natexlab{}.
\newblock \showarticletitle{Implementing programming in school mathematics and
  technology: teachers’ intrinsic and extrinsic challenges}.
\newblock \bibinfo{journal}{\emph{International journal of technology and
  design education}} (\bibinfo{year}{2020}), \bibinfo{pages}{1--30}.
\newblock


\bibitem[\protect\citeauthoryear{Yadav, Gretter, Hambrusch, and Sands}{Yadav
  et~al\mbox{.}}{2016}]%
        {yadav2016expanding}
\bibfield{author}{\bibinfo{person}{A. Yadav}, \bibinfo{person}{S. Gretter},
  \bibinfo{person}{S. Hambrusch}, {and} \bibinfo{person}{P. Sands}.}
  \bibinfo{year}{2016}\natexlab{}.
\newblock \showarticletitle{Expanding computer science education in schools:
  understanding teacher experiences and challenges}.
\newblock \bibinfo{journal}{\emph{Computer Science Education}}
  \bibinfo{volume}{26}, \bibinfo{number}{4} (\bibinfo{year}{2016}),
  \bibinfo{pages}{235--254}.
\newblock


\bibitem[\protect\citeauthoryear{Yeni, Aivaloglou, and Hermans}{Yeni
  et~al\mbox{.}}{2020}]%
        {yeni2020or}
\bibfield{author}{\bibinfo{person}{S. Yeni}, \bibinfo{person}{E. Aivaloglou},
  {and} \bibinfo{person}{F. Hermans}.} \bibinfo{year}{2020}\natexlab{}.
\newblock \showarticletitle{To Be or Not to Be a Teacher? Exploring CS
  Students’ Perceptions of a Teaching Career}. In
  \bibinfo{booktitle}{\emph{Koli Calling}}. \bibinfo{pages}{1--11}.
\newblock


\bibitem[\protect\citeauthoryear{Yukselturk and Altiok}{Yukselturk and
  Altiok}{2017}]%
        {yukselturk2017investigation}
\bibfield{author}{\bibinfo{person}{E. Yukselturk} {and} \bibinfo{person}{S.
  Altiok}.} \bibinfo{year}{2017}\natexlab{}.
\newblock \showarticletitle{An investigation of the effects of programming with
  Scratch on the preservice IT teachers’ self-efficacy perceptions and
  attitudes towards computer programming}.
\newblock \bibinfo{journal}{\emph{BJET}} \bibinfo{volume}{48},
  \bibinfo{number}{3} (\bibinfo{year}{2017}), \bibinfo{pages}{789--801}.
\newblock


\bibitem[\protect\citeauthoryear{Zhan, Fong, Mei, and Liang}{Zhan
  et~al\mbox{.}}{2015}]%
        {zhan2015effects}
\bibfield{author}{\bibinfo{person}{Z. Zhan}, \bibinfo{person}{P.~S. Fong},
  \bibinfo{person}{H. Mei}, {and} \bibinfo{person}{T. Liang}.}
  \bibinfo{year}{2015}\natexlab{}.
\newblock \showarticletitle{Effects of gender grouping on students’ group
  performance, individual achievements and attitudes in computer-supported
  collaborative learning}.
\newblock \bibinfo{journal}{\emph{Comput. Hum. Behav.}}  \bibinfo{volume}{48}
  (\bibinfo{year}{2015}), \bibinfo{pages}{587--596}.
\newblock


\end{thebibliography}

\end{document}